\documentclass[11pt]{article}
\pdfoutput=1
\usepackage{color}
\usepackage{amsmath,amssymb,amsfonts,dsfont}
\usepackage{graphicx}
\usepackage{subfigure}
\usepackage[citebordercolor={.8 .8 1},urlbordercolor ={.8 .8 1},pdfstartview=FitV]{hyperref}

\newcommand{\beq}{\begin{equation}}
\newcommand{\eeq}{\end{equation}}
\newcommand{\beqn}{\begin{eqnarray}}
\newcommand{\eeqn}{\end{eqnarray}}

\newcommand{\gc}{q} 

\numberwithin{equation}{section}

\begin{document}
\begin{center}

\vspace{1cm} { \LARGE {\bf Boundary Conditions and Unitarity: the Maxwell-Chern-Simons System in AdS$_{3}$/CFT$_{2}$}}

\vspace{1.1cm}
Tom\'as Andrade$^{1}$, Juan I. Jottar$^{2,3}$ and Robert G. Leigh$^{2,4}$

\vspace{0.7cm}

{\it $^{1}$Department of Physics, UCSB\\
  Santa Barbara, CA 93106, U.S.A.}

\vspace{0.5cm}

{\it $^{2}$Department of Physics, University of Illinois,\\
      1110 W. Green Street, Urbana, IL 61801, U.S.A. }

\vspace{0.5cm}

{\it $^{3}$Institute for Theoretical Physics, University of Amsterdam,\\
Science Park 904, Postbus 94485, 1090 GL Amsterdam, The Netherlands}

\vspace{0.5cm}

{\it $^{4}$Perimeter Institute for Theoretical Physics,\\ 31 Caroline St. N., Waterloo, ON, Canada N2L 2Y5 }

\vspace{0.5cm}

{\tt tandrade@umail.ucsb.edu, J.I.Jottar@uva.nl, rgleigh@illinois.edu} \\

\vspace{1.5cm}

\end{center}

\begin{abstract}
\noindent We consider the holography of the Abelian Maxwell-Chern-Simons (MCS) system in Lorentzian three-dimensional asymptotically-AdS spacetimes, and discuss a broad class of boundary conditions consistent with conservation of the symplectic structure. As is well-known, the MCS theory contains a massive sector dual to a vector operator in the boundary theory, and a topological sector consisting of flat connections dual to $U(1)$ chiral currents; the boundary conditions we examine include double-trace deformations in these two sectors, as well as a class of boundary conditions that mix the vector operators with the chiral currents.
We carefully study the symplectic product of bulk modes and show that almost all such boundary conditions induce instabilities and/or ghost excitations, consistent with violations of unitarity bounds in the dual theory.

\vspace{30pt}
\end{abstract}

\pagebreak

\setcounter{page}{1}
\setcounter{equation}{0}
\tableofcontents

\pagebreak

\section{Introduction}
Since the early days of the holographic correspondence \cite{Maldacena:1997re,Witten:1998qj,Gubser:1998bc}, $(2+1)$-dimensional gravitational systems have played a central role in testing and exploring the ideas behind the duality. In fact, with the benefit of hindsight, one can see that the work of Brown and Henneaux on the asymptotic symmetries of three-dimensional spacetimes with a negative cosmological constant \cite{Brown:1986nw} displayed some basic features of the correspondence as early as a decade before it was proposed. The 3$d$ gravitational systems of interest in the framework of holography are special in that their field theory duals enjoy an infinite-dimensional (local) conformal symmetry; via the powerful techniques of conformal field theory (CFT), one then has a better grasp of the boundary theory structure which is often lacking in higher-dimensional examples. A beautiful example of this fact is the precise connection between the CFT spectrum and retarded Green's functions, and black hole quasinormal modes in the bulk \cite{Birmingham:2001pj}.

The feature that makes the gauge/gravity correspondence outstanding is that it postulates the equivalence between gravitational weakly-coupled degrees of freedom propagating in the bulk spacetime, and strongly-coupled degrees of freedom in a dual quantum field theory in one less dimension (``the boundary"). A pivotal ingredient in the proposal is the dilatation symmetry of the boundary CFT; broadly speaking, using this symmetry one can relate the masses of the bulk fields to the conformal dimensions of operators in the quantum theory on the boundary, as first established in \cite{Witten:1998qj}. However, for a given bulk field, the spectrum of conformal dimensions in the dual quantum field theory is not entirely determined by the masses of the fields. This is intimately related to the fact that the boundary conditions that yield well-defined dynamics are not unique. In fact, in the so-called ``bottom-up" holography where the bulk theory is phenomenologically devised, the operator content of the possible dual theories is completely specified only after the boundary conditions for bulk fields have been properly chosen.

In the present article, we will focus on the study of the Abelian Maxwell-Chern-Simons (MCS) theory, frequently referred to as ``Topologically Massive Electrodynamics" \cite{Deser:1981wh,Deser:1982vy}, in three-dimensional asymptotically-AdS spacetimes. The emphasis will be on determining a set of ``admissible" boundary conditions, in a sense that will be made precise below; this is crucial to the dictionary problem in the context of the AdS/CFT correspondence, as discussed above. One of the motivations to study Chern-Simons terms in the bulk is that these arise naturally in the context of string theory compactifications, and endow the bulk black hole solutions with $U(1)$ charge (see \cite{Kraus:2006nb,Kraus:2006wn}, for example). It is worth mentioning, however, that the MCS system plays a central role in condensed matter physics as well, in particular in the study of fermionic systems in two spatial dimensions, where it describes the low-energy effective theory of the Fractional Quantum Hall Effect (FQHE). Furthermore, even in flat space the MCS theory is often said to be holographic, albeit in a different sense from the above: in the topological limit (where the bulk quasiparticles become infinitely massive), the degrees of freedom are effectively localized on the boundary \cite{Wen:1991mw}.\footnote{The key difference being that the AdS/CFT correspondence is an \textit{equivalence} between bulk and boundary degrees of freedom, while in the topological theory the bulk degrees of freedom are gapped, and the low-energy excitations propagate exclusively on the boundary.} From a mathematical point of view, this is the well-known correspondence between three-dimensional Chern-Simons gauge theory and a chiral rational CFT \cite{Witten:1988hf, Moore:1989yh, Elitzur:1989nr,Balachandran:1991dw, Bos:1989kn, Bos:1989wa,Schwarz:1979ae}. More recently, the latter correspondence has been refined to reconcile the modular transformation properties of the string theory partition function on $AdS_3$ and those of the Chern-Simons theory which dominates its infrared dynamics \cite{Gukov:2004id}, and the potential relevance of these observations for condensed matter physics was also pointed out. This provides yet another motivation to carefully study the holographic dictionary of the full (finite coupling) MCS theory; here we will do so from a bottom-up perspective, in the hope that our results could be useful in the study of other models which might be interesting for applications of holography to condensed matter physics.

Our analysis starts by determining a broad set of boundary conditions under which the bulk theory is expected to have well-posed dynamics. We find it convenient to approach this problem using the covariant phase space formalism, along the lines of \cite{Marolf:2006nd,Amsel:2006uf,Amsel:2008iz}. Within this framework, the first requirement on the boundary conditions is that they lead to a conserved symplectic structure (in the sense of timelike evolution). In the context of holography, this condition can be conveniently rephrased as the vanishing of the symplectic flux on the (radial) boundary. Roughly speaking, the bulk gauge field splits into a ``massive", gauge-invariant piece, and the flat connections. Accordingly, the boundary theory operators organize themselves into two sectors: a vector operator dual to the massive part of the connection, and the well-known $U(1)$ chiral currents (which also arise in the pure Chern-Simons theory). We will obtain a variety of boundary conditions that correspond to double-trace deformations from the dual field theory perspective. In particular, we shall note the possibility of coupling the vector operator and the chiral currents via this mechanism. To our knowledge, these ``hybrid" boundary conditions intertwining the massive and topological sectors have not been discussed in the literature; their existence was anticipated in \cite{Gukov:2004id}, however, where the topology of the spacetime manifold was chosen in such a way that the two sectors effectively decouple.

It is worth emphasizing that all of this physics occurs at finite Maxwell coupling. It is often argued in the literature that the Maxwell coupling should be irrelevant in the infrared; this is certainly true from the bulk perspective. However, the Maxwell coupling is not irrelevant in the UV, and so is an important parameter holographically. One also notes in parallel that in condensed matter systems such as quantum Hall, the Maxwell coupling sets the cut-off scale where quasi-particle excitations live, going away only in the topological limit. It seems quite plausible that such excitations will exist in the holographic theory as well, a subject that we will explore elsewhere.

Having obtained the class of boundary conditions that lead to a conserved symplectic structure, one can examine in detail which of these are consistent with unitarity. Our motivation to consider this restriction comes primarily from the existence of the unitarity bound in conformal field theories (see \cite{Mack:1975je,Minwalla:1997ka}, for example), which dictates that the presence of operators whose dimension is ``too low" leads to negative norm states (ghosts). Via the holographic correspondence, this fact should manifest in the bulk physics as well, which is the question we address. As first noted in \cite{Balasubramanian:1998sn,Witten:2001ua}, a closely analogous concern arises when considering bulk scalar fields with sufficiently high masses if one imposes boundary conditions that allow the slow-decaying branch to fluctuate. As a result of this choice, the conformal dimension of the dual operator lies below the unitarity bound and one expects the bulk theories to be ill-defined. Recently, these setups were considered in \cite{Andrade:2011dg}, which confirmed that such bulk theories are indeed pathological and that, generically, they suffer from ghosts.

 We will address the question of unitarity by studying the dynamics of the MCS system in $AdS_{3}$ in both global and Poincar\'e coordinates; in particular, we will discuss the resulting spectrum and symplectic products for the various boundary conditions for which the symplectic structure is conserved. Our main result is that the only boundary conditions consistent with unitarity do not mix the massive and topological sectors, and in particular they require to hold fixed the slower fall-off of the massive mode (i.e. they are of Dirichlet type). In short, the class of permissible boundary conditions is severely restricted by unitarity considerations. Interestingly, we will also find additional ghosts in the flat sector whose presence cannot be linked to unitarity bounds in an obvious way. We will also include an analysis of the symmetries we expect to be present in the dual theory as a result of various choices of boundary conditions.

The three-dimensional MCS theory has been previously considered in the context of AdS/CFT. We refer the reader to \cite{Gukov:2004id,Kraus:2006wn, Kraus:2006nb,Jensen:2010em} for work which focuses on the flat (topological) sector of the theory. The massive sector has also received some attention and the holographic dictionary problem has been studied to some extent \cite{Minces:1999tp,DHoker:2010hr,Yee:2011yn}. Our results agree with the references above as far as the operator content is concerned. The novelty of our analysis lies in the fact that we have considered a wider class of boundary conditions, including ``hybrid" boundary conditions that mix the massive and topological sectors, and analyzed their consistency with unitarity in detail. Additional related work includes \cite{Carlip:2008jk,Fujita:2009kw,Balasubramanian:2010sc,Fosco:2011ra}.

This paper is organized as follows. In section \ref{sec: prelim} we review the MCS system, the solution of the asymptotic equations of motion on asymptotically $AdS_3$ backgrounds, and the corresponding conformal dimensions of dual operators. In section \ref{section:symplectic} we briefly describe the covariant phase space formalism, and use the conservation of the symplectic structure as a criterion to determine a wide class of \textit{a priori} admissible boundary conditions in the holographic setup. For all the boundary conditions of interest, we construct the appropriate action principles and compute the one-point functions of the dual operators holographically. We also review the notion of symplectic product, which will play a central role in the analysis of unitarity. In section \ref{section:spectrum} we discuss the spectrum of excitations in the dual field theory for the class of boundary conditions previously found, and discuss the normalizability of the various bulk modes. In section \ref{section:norms} we present the calculation of the symplectic product for the various normalizable modes, focusing on the existence of ghosts; the requirement of unitarity in the dual theory then leads to a restricted class of permissible boundary conditions, which constitutes our main result. We conclude in section \ref{section:discussion} with a discussion of our findings, along with possible extensions and applications. Some useful results used in the body of the paper have been collected in the appendices, as well as a brief discussion of the $U(1)$ symmetries in the dual field theory for the different boundary conditions under consideration.

\section{The Maxwell-Chern-Simons system}\label{sec: prelim}
We consider the Maxwell-Chern-Simons (MCS) system in $(2+1)$ spacetime dimensions,
\begin{align}\label{TME bulk action}
 I
={}&
- \frac{1}{4\gc^2}\int_{M}d^{3}x\,\sqrt{|g|}F_{\mu\nu}F^{\mu\nu} -\frac{\hat{\alpha}}{4} \int_{M}d^{3}x\, \varepsilon^{\mu\nu\rho}A_{\mu}F_{\nu\rho}
 \, ,
\end{align}
\noindent where $\gc^2$ is the gauge coupling (with units of $[length^{-1}]$) and $\hat{\alpha}$ is the (dimensionless) Chern-Simons (CS) coupling. Throughout this paper we work in a fixed background in which we neglect the backreaction of the gauge field on the metric, {\it i.e.} $G_N/\gc^2 \to 0$, where  $G_N$ is the three dimensional gravitational coupling (which has units of $[length]$). Where appropriate, we will occasionally comment on issues of backreaction, and will consider them in a subsequent publication. The background metrics we consider are solutions of the Einstein equations in the presence of a negative cosmological constant $\Lambda = -1/L^2$, and the normalization is chosen such that pure $AdS_3$ space is a vacuum solution of the decoupled gravitational sector with radius $L$ and scalar curvature $R=-6/L^2$. As usual, in a holographic context the action \eqref{TME bulk action} must be supplemented by a collection of boundary terms that render the variational problem well defined and remove divergent contributions; these will be fully specified later on in the paper.

The equations of motion that follow from \eqref{TME bulk action} are\footnote{Our convention for the Levi-Civita tensor is $\epsilon^{\mu\nu\rho} = -\frac{1}{\sqrt{|g|}}\varepsilon^{\mu\nu\rho}$, where $\varepsilon^{\mu\nu\rho}$ is the Levi-Civita symbol.}
\begin{align}
\nabla_{\nu}F^{\nu\mu} + \frac{\alpha}{2L}\epsilon^{\mu\nu\rho}F_{\nu\rho}=0\, ,\label{MaxCS equation}
\end{align}
\noindent where we have defined the rescaled CS coupling $\alpha$ as
\begin{equation}\label{rescaled CS coupling}
\alpha = \gc^2 L\hat{\alpha}\, ,
\end{equation}
\noindent which is also dimensionless. Without loss of generality, we will assume $\alpha > 0$. When taking backreaction on the metric into account, asymptotically AdS solutions exist only for $\alpha<1$, and we will restrict our discussions in the present paper to that range.

In form language, the Maxwell-CS equation \eqref{MaxCS equation} can be written as\footnote{On a $D$-dimensional spacetime, our convention for the Hodge dual is $*(dx^{\nu_{1}}\wedge\cdots \wedge dx^{\nu_{r}}) = \frac{1}{(D-r)!}\epsilon^{\nu_{1}\ldots \nu_{r}}_{\phantom{\nu_{1}\ldots \nu_{r}}\mu_{1}\ldots \mu_{D-r}}dx^{\mu_{1}}\wedge\cdots \wedge dx^{\mu_{D-r}}$.}
\begin{equation}\label{Maxwell-CS eom}
d^{\dagger}F = \frac{\alpha}{L}*F
\end{equation}
\noindent where $d^\dagger$ is the adjoint exterior derivative, which in our conventions acts on $F$ as $d^\dagger F = -*d(*F) = -\nabla_{\mu}F^{\mu}_{\phantom{\mu}\nu}\, dx^{\nu}$. Hence, the equation of motion implies
\begin{equation}\label{gauge field splitting}
    A = A^{(0)} + B\, ,
\end{equation}
\noindent  where $A^{(0)}$ is a flat connection and we have defined
\begin{equation}\label{definition B}
B \equiv  -\frac{L}{\alpha}*F\,.
\end{equation}
\noindent We note that $B$ is, by definition, invariant under the $U(1)$ gauge symmetry of the theory. In a later section we will study the consequences of the splitting \eqref{gauge field splitting} at the level of the symplectic structure and the boundary conditions in a holographic context.

Since $dB=dA=F$, the equation of motion \eqref{Maxwell-CS eom} becomes a first order equation for $B$:
\begin{equation}\label{equation of motion for B}
*dB + \frac{\alpha}{L}B=0\, ,
\end{equation}
\noindent which is the familiar equation for a massive vector field. In components, this equation reads
\begin{equation}\label{first order equation for B}
 \epsilon^{\mu\nu\rho}\partial_{\nu}B_{\rho} + \frac{\alpha}{L}B^{\mu}=0\, .
\end{equation}
\noindent Notice also that the definition \eqref{definition B} implies a consistency condition:
\begin{equation}
d^\dagger B = 0\, ,
\end{equation}
\noindent i.e. $B$ is a co-closed form ($\nabla^{\mu}B_{\mu}=0$); naturally, this also follows from the equation of motion \eqref{equation of motion for B}.  Acting on \eqref{equation of motion for B} with $*d$ we can write a second-order equation for $B$,
\begin{equation}
0 = d^{\dagger}d B +\frac{\alpha^2}{L^2}B=\Delta B + \frac{\alpha^2}{L^2}B\, ,
\end{equation}
\noindent where $\Delta = d^\dagger d + d d^\dagger$ is the Laplacian.

\subsection{Asymptotic solutions}\label{subsection:asymptotic sols}
For the sake of concreteness,  we will write the metric of the asymptotically AdS spacetimes we are interested in  as
\begin{equation}\label{asympt expansion of metric}
ds^{2} \xrightarrow[r\to \infty]{} L^{2}\frac{dr^{2}}{r^{2}} + \frac{r^{2}}{L^{2}}g^{(0)}_{ij}(x)dx^{i}dx^{j} + \ldots
\end{equation}
\noindent Restricting ourselves to flat connections which are finite at the conformal boundary, the asymptotic form of the solution for the gauge field is then of the form\footnote{As we will see in appendix \ref{section:sym bndy}, any finite $r$-dependent piece in the near-boundary behavior of the flat connection can be removed with the appropriate gauge transformation.}
\begin{equation}\label{gen asympt}
A(r,x) \xrightarrow[r\to \infty]{} A^{(0)}(x) + r^{\alpha}\left(B^{(+)}(x) +\mathcal{O}(r^{-2})\right)+ r^{-\alpha}\left(B^{(-)}(x)+\mathcal{O}(r^{-2})\right)  \,,
\end{equation}
where $A^{(0)}$ is flat, i.e. $F^{(0)}=dA^{(0)}=0$. Similarly, solving the equations of motion asymptotically  one finds that the radial component $B_{r}$ of the gauge-invariant mode is subleading with respect to the $B_{i}$ components, which are moreover constrained by
\begin{equation}\label{asymptotic constraint}
P_{\pm}^{ij}B_{j}^{\left(\mp\right)} = 0\, ,\qquad \mbox{where } \quad  P_{\pm}^{ij} = \frac{1}{2}\left(g^{(0) ij} \pm  \epsilon^{ij}\right) .
\end{equation}
\noindent We have adopted the convention that $ \epsilon^{ij} = -\varepsilon^{ij}/\sqrt{|g^{(0)}|}$, where $\varepsilon^{ij}$ is the two-dimensional Levi-Civita symbol, related to its three-dimensional counterpart by $\varepsilon^{ij} = \varepsilon^{rij}$. Notice that the projectors $P_{\pm}^{ij}$ satisfy the usual properties: $\left(P_{+}P_{-}\right)^{ij} = P_{+}^{ik}g^{(0)}_{kl}P_{-}^{lj}=0$, $\left(P_{\pm}^{2}\right)^{ij} = P_{\pm}^{ik}g^{(0)}_{kl}P_{\pm}^{lj} = P_{\pm}^{ij}\,$.

\subsection{Conformal dimensions}\label{subsection:conf dims}
Given the asymptotic expansion \eqref{gen asympt} and noting that the pullback to the boundary of the bulk vector field is simply a boundary vector, we conclude that the standard AdS/CFT dictionary relates $B^{(+)}$ and $B^{(-)}$ with vector operators of dimensions $\Delta_- = 1 - \alpha$ and $\Delta_+ = 1 + \alpha$, respectively. On the other hand, the components of $A^{(0)}$ have scaling dimension one. As we shall review below, the components of $A^{(0)}$ along the boundary directions correspond to chiral currents that live on the boundary theory, \cite{Witten:1988hf, Moore:1989yh, Elitzur:1989nr,Wen:1991mw,Balachandran:1991dw, Bos:1989kn, Bos:1989wa,Schwarz:1979ae}. We note that the lower scaling dimension is positive as long as $\alpha < 1$, which implies that we can allow both fall-offs to fluctuate while preserving locally AdS asymptotics\footnote{Here we use the terminology of \cite{Skenderis:2002wp}, i.e., we mean that the curvature near the conformal boundary is that of AdS plus subleading corrections.} if $\alpha < 1$. We have verified this statement explicitly by studying the effect of backreaction on a general asymptotically locally AdS metric of the form \eqref{asympt expansion of metric}.

It should be noted that the operator of dimension $\Delta_-$ violates the unitarity bound $\Delta_V = 1$ for vector operators in two dimensions for all $\alpha>0$
\cite{Mack:1975je,Minwalla:1997ka}, see also \cite{ElShowk:2011gz} for the explicit expression. This suggests that boundary conditions that allow this degree of freedom to fluctuate should yield pathologies in the bulk; in subsequent sections we shall verify that this is indeed the case.

\section{Symplectic structure and boundary conditions}\label{section:symplectic}
In the present section we study the issue of boundary conditions in the holographic description of the MCS system. We find it convenient to work within the covariant phase space formalism, which we will review shortly. The motivation for employing this formalism is two-fold: first, the classification of the allowed boundary conditions is nicely encoded in a simple vanishing-flux condition; and second, it allows us to classify the spectrum of excitations in a clean way. We emphasize however that this decision is just a matter of personal preference, and the results obtained within this framework should indeed be equivalent to the ones arrived at by more familiar, say canonical, methods.

We now proceed to briefly review the covariant phase space techniques; more detailed discussions can be found in \cite{Lee:1990nz, Wald:1995yp,Wald:1999wa,Iyer:1994ys,Ashtekar:1990gc}. First, we stress that the construction is inherently Lorentzian, so we shall assume that the spacetime is endowed with a Lorentzian metric. Now, the ingredient that lies at the heart of this construction is the identification of the phase space with the space of solutions of the equations of motion which satisfy certain boundary conditions. This is possible since in any well-defined setup the specification of a point in canonical phase-space, i.e. of initial data, completely determines the subsequent evolution of the system. The other main ingredient is an algebraic structure that determines the dynamics once a Hamiltonian function is given, or crudely speaking, something that contains information about the Poisson brackets. This is nothing but the pre-symplectic structure of the theory, $\Omega$, which can be thought of as a (possibly degenerate) two-form in the tangent space of (linearized) solutions. In other words, $\Omega$ maps a pair of tangent vectors in the space of solutions to the real numbers. Given a background solution $\bar{s}$ and two linearized solutions $\delta_1 s$ and $\delta_2 s$, we denote the symplectic product of $\delta_1 s$ with $\delta_2 s$ by $\Omega(\delta_1 s, \delta_2 s; \bar{s})$. Quite conveniently, this object can be constructed algorithmically given a Lagrangian \cite{Lee:1990nz}, and we will illustrate this below.

From the discussion above, it follows that the pre-symplectic structure must indeed be conserved in order for the identification of the initial data with the space of solutions to be independent of the surface on which the initial data is specified. This conservation condition is what we shall take as a guiding principle to classify the allowed boundary conditions for the MCS system. It is worth emphasizing here that the boundary conditions are in fact a crucial part of the definition of the phase space of a given theory. As pointed out above, the covariant phase space formalism also provides a useful way to classify the spectrum of excitations of the theory. In particular, we mention that in the presence of gauge symmetries the pre-symplectic structure is degenerate, the gauge orbits being precisely its null directions. Thus, we shall refer to any solution of the equations of motion whose symplectic product with an arbitrary solution vanishes as ``pure gauge".\footnote{We mention that the prefix ``pre" makes reference to the degeneracy of $\Omega$: by definition, a symplectic structure is non-degenerate. In a slight  abuse of notation we drop the prefix from now on, even when the kernel of $\Omega$ is non-empty.} Further nomenclature will be discussed in section \ref{sec: symp prod}.

After taking the quotient by the gauge directions, the symplectic structure has a unique inverse and this corresponds to the Poisson bracket defined for gauge-invariant quantities. As discussed in detail in \cite{Wald:1995yp}, this relation can be written succinctly as
\begin{equation}\label{PB omega}
    \{ \Omega(\delta_1 s, \cdot; \bar{s}) , \Omega(\delta_2 s, \cdot; \bar{s}) \}_{PB} = - \Omega(\delta_1 s, \delta_2 s; \bar{s}) \, .
\end{equation}
\noindent Here $\Omega(\delta_1 s, \cdot; \bar{s})$ is to be understood as a linear function in covariant phase space. Then, the fact that the Poisson bracket and $\Omega$ are the inverse of each other follows trivially by writing \eqref{PB omega} in component notation. Finally, we mention that, at the classical level, one can construct conserved charges directly in terms of $\Omega$. More precisely, given an infinitesimal transformation $\delta_\lambda s$ and an arbitrary linearized solution $\delta s$, the infinitesimal variation of the generator $Q_\lambda$ along $\delta s$ is given by
\begin{equation}\label{d Q}
    \delta Q_\lambda  = \Omega(\delta_\lambda s, \delta s;\bar{s}) \, ,
\end{equation}
\noindent which, once again, is most easily visualized by translating \eqref{d Q} into component notation. We stress that the charge $Q_\lambda$ is only defined if \eqref{d Q} is finite and satisfies the appropriate integrability conditions, see e.g. \cite{Wald:1993nt}. Expression \eqref{d Q} also makes it clear that gauge transformations, i.e. null directions of $\Omega$, have a vanishing generator. This is just the familiar statement that the generators of gauge symmetries are constraints, and as such vanish on-shell.  On the other hand, global symmetries are associated to a non-zero charge.

\subsection{The symplectic flux}\label{subsection:symplectic flux}
In this section we apply the method of \cite{Lee:1990nz} to construct the symplectic structure of the MCS theory and determine the expression for the symplectic flux, which serves as a first step in classifying the allowed boundary conditions. Under an infinitesimal variation $\delta A_{\mu}$ of the gauge field  (and assuming a fixed background metric), the first order variation of the bulk action\footnote{Note that we have not included the boundary terms in the action here. We will come back to them later, and confirm that they do not contribute to the symplectic structure.} is
\begin{equation}\label{first variation action}
\delta I = \int _{M}d^{3}x\sqrt{|g|}\, \mbox{EOM}(A)^{\mu}\delta A_{\mu} - \int_{\partial M}d^{2}x\,\sqrt{|\gamma|}\, \rho_{\mu}\left(\frac{1}{q^2}F^{\mu\nu} + \frac{\hat{\alpha}}{2}\epsilon^{\mu\rho\nu}A_{\rho}\right)\delta A_{\nu}\, ,
\end{equation}
\noindent where $\mbox{EOM}(A)^{\mu}=0$ is the equation of motion of the background gauge field, $\gamma$ is the determinant of the induced metric on the timelike boundary (a ``constant radius" slice), and $\rho^{\mu}$ denotes the corresponding unit normal.  From the above variation we read off the symplectic 1-form (see \cite{Lee:1990nz,Compere:2008us})
\begin{align}
\theta^{\mu}
={}&
 - \left(\frac{1}{q^2}F^{\mu\nu} + \frac{\hat{\alpha}}{2}\epsilon^{\mu\rho\nu}A_{\rho}\right)\delta A_{\nu}\, .
\end{align}
\noindent Next, denoting by $\delta_{1}A$ and $\delta_{2}A$ two independent solutions of the linearized equations of motion\footnote{We note that in the probe approximation the equations of motion for the background gauge field and its fluctuation have the same form, because the MCS system is linear.} we define the symplectic 2-form
\begin{align}
\omega^{\mu}(\delta_1 A, \delta_2 A; \bar{A})
\equiv{}&
	\delta_{1}\theta[\delta_{2}A] - \delta_{2}\theta[\delta_{1}A]
	\nonumber\\
={}&
	-\frac{1}{q^2}\Bigl(\delta_{1}F^{\mu\nu}\delta_{2} A_{\nu}-\delta_{2}F^{\mu\nu}\delta_{1} A_{\nu}\Bigr) -\hat{\alpha}\, \epsilon^{\mu\rho\nu}\delta_{1}A_{\rho}\delta_{2} A_{\nu}\, .
\end{align}
\noindent Using the equation of motion for $\delta F_{\mu\nu}$ (which is the same as \eqref{MaxCS equation}, because we are ignoring backreaction on the metric) one can then show the crucial property
\begin{equation}\label{symplectic continuity eq}
\nabla_{\mu}\omega^{\mu} = 0\, .
\end{equation}

As stated above, we assume that the $(2+1)$ manifold is Lorentzian, with the topology $X\times \mathds{R}$, where the $\mathds{R}$ factor is parameterized by the timelike coordinate ($t$, say). The boundary $\partial M$ is a surface of constant $r$. We now define the symplectic structure by
\begin{equation}\label{OM bulk}
\Omega(\delta_1 A, \delta_2 A; \bar{A}) = \int_{\Sigma}d^{2}x\sqrt{h }\, n_{\mu}\omega^{\mu}\, ,
\end{equation}
\noindent where $\Sigma$ is a spacelike hypersurface (a $t={\rm constant}$ slice, for example) with unit normal $n^{\mu}$ and induced metric determinant $h$. Since the theory under consideration is linear, we can take the background to be the trivial configuration, i.e. $\bar{A} = 0$, without loss of generality. We shall do so henceforth and omit the explicit reference to the background as an argument of the symplectic structure. We mention that, in principle, the bulk expression \eqref{OM bulk} may require renormalization; the appropriate counterterms can be read off from a well-defined action principle as explained in \cite{Compere:2008us}. However, working in the range $0 < \alpha < 1$, no (UV) divergences arise in \eqref{OM bulk} even if we allow the slow fall-off of the field to fluctuate, as we will verify by explicit computation in section \ref{section:norms}. This is intimately related to the fact that, for $0 < \alpha < 1$, the counterterms that render the variational principle well-defined do not include derivatives along the timelike direction, see section \ref{sec:1pt}.

\begin{figure}[htb]
\center
\includegraphics[width=0.65\linewidth]{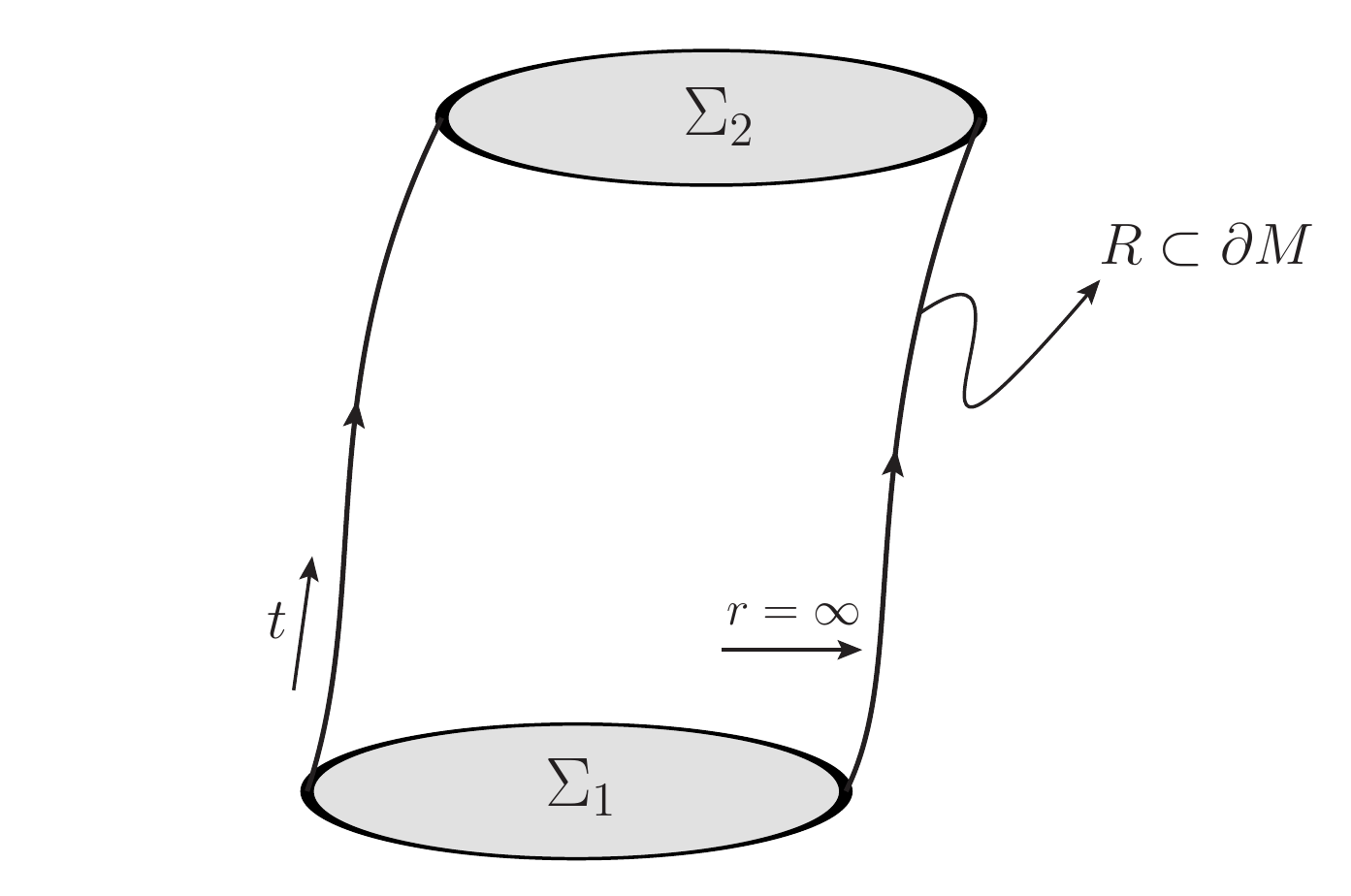}
\label{pillbox}
\caption{The symplectic structure is conserved, i.e. $\Omega(\Sigma_{1})=\Omega(\Sigma_{2})$, when the symplectic flux through the region $R \subset \partial M$ vanishes.}
\end{figure}

As discussed above, in order to obtain a well-defined phase space it is necessary to impose boundary conditions on our solutions in such a way that the symplectic structure is \textit{conserved} (i.e. independent of $\Sigma$). Integrating equation \eqref{symplectic continuity eq} over a ``pillbox" bounded by two spacelike hypersurfaces $\Sigma_{1}$ and $\Sigma_{2}$ and a region $R \subset \partial M$ (i.e. $R$ is an open subset of the boundary slice at constant $r$, see figure \ref{pillbox}), one learns that the symplectic structure is independent of $\Sigma$ provided the symplectic flux $\Phi$ through $R$ vanishes, i.e.
\begin{equation}\label{vanishing of flux}
\Phi = \int_{R}d^{2}x\sqrt{|\gamma|}\,\rho_{\mu}\omega^{\mu} =0\, ,
\end{equation}
\noindent where, as before, $\rho^{\mu}$ and $\gamma$ are the unit normal and the determinant of the induced metric on $R$, respectively. We suppose that this is attained {\it locally}, so that the flux through the boundary vanishes through any open subset $R$. We mention that, from the point of view of the dual theory, these local boundary conditions correspond to the insertion of local operators. In the presence of additional boundaries, e.g. the Poincar\'e horizon, one must also require the flux to vanish there. Given our assumption of locality, the boundary conditions at the extra boundaries are of course independent of the ones at the conformal boundary. It is worth noting that, for black hole spacetimes, the phase space is typically defined including the interior of the black hole, so a non-vanishing flux through the horizon is not in conflict with conservation of $\Omega$.

In the coordinates introduced in \eqref{asympt expansion of metric} the only non-vanishing component of $\rho$ is $\rho_{r} = \sqrt{g_{rr}} = N_{r}$, where $N_{r}$ is the lapse in a radial foliation. Since $\sqrt{|g|} = N_{r}\sqrt{|\gamma|}$, we have
\begin{equation}
\Phi  = \int_{R}d^{2}x\sqrt{|g|}\,\bar{\rho}_{\mu}\omega^{\mu}\, ,
\end{equation}
\noindent where $\bar{\rho}_{\mu}\, dx^{\mu}=dr$ and $g$ is the determinant of the full $(2+1)$ metric, as before. If we now split the connection as in \eqref{definition B}, so that in an obvious notation the gauge field fluctuation is $\delta A  = \delta B + \delta A^{(0)}$, we find
\begin{equation}
\omega^{\mu} = \omega^{\mu}_{B} + \omega^{\mu}_{0} + \omega^{\mu}_{mix}\, ,
\end{equation}
\noindent where we have defined
\begin{align}
\omega^{\mu}_{B}
\equiv{}&
	-\frac{1}{q^2}\Bigl(f_{1}^{\mu\nu}\delta_2 B_{\nu}-f_{2}^{\mu\nu}\delta_1 B_{\nu}\Bigr) -\hat{\alpha}\, \epsilon^{\mu\rho\nu}\delta_1 B_{\rho} \delta_2 B_{\nu}
\\
\omega^{\mu}_{0}
\equiv {}&
	-\hat{\alpha}\, \epsilon^{\mu\rho\nu}\delta_1 A^{(0)}_{\rho} \delta_2 A^{(0)}_{\nu}
\\
\omega^{\mu}_{mix}
\equiv {}&
	-\frac{1}{q^2}\Bigl(f_{1}^{\mu\nu}\delta_2 A^{(0)}_{\nu}-f_{2}^{\mu\nu}\delta_1 A^{(0)}_{\nu}\Bigr) -\hat{\alpha}\, \epsilon^{\mu\rho\nu}\bigl(\delta_1 A^{(0)}_{\rho} \delta_2 B_{\nu} + \delta_1 B_{\rho} \delta_2 A^{(0)}_{\nu}\bigr)
\end{align}
\noindent with $f$ the field strength of $\delta B$. We now notice that contracting equation \eqref{first order equation for B} with the Levi-Civita tensor results in $0 = F_{\mu\nu} - q^{2}\hat{\alpha}\,\epsilon_{\mu\nu\rho}B^{\rho}$. Consequently, the fluctuations of the gauge-invariant mode satisfy
\begin{equation}\label{massive mode fluctuation eq}
f^{\mu\nu} =   q^{2}\hat{\alpha}\,\epsilon^{\mu\nu\rho}\delta B_{\rho}\, .
\end{equation}

\noindent Using this on-shell condition in the above expression for $\omega^{\mu}$ we find
\begin{align}\label{omegas flat and non-flat}
\omega^{\mu}_{B}
=
	\hat{\alpha}\,\epsilon^{\mu\nu\rho}\delta_1 B_{\nu}\delta_2 B_{\rho}\, ,\qquad
\omega^{\mu}_{0}
=
	-\hat{\alpha}\, \epsilon^{\mu\nu\rho}\delta_1 A^{(0)}_{\nu} \delta_2 A^{(0)}_{\rho}
\,, \qquad
\omega^{\mu}_{mix}
=
	0\, .
\end{align}

As a result of the splitting \eqref{omegas flat and non-flat}, the symplectic structure can be written as
\begin{equation}\label{omega split}
    \Omega = \int_\Sigma d^2 x \sqrt{h}\, n_\mu \omega^{\mu}_{B} + \int_\Sigma d^2 x \sqrt{h}\, n_\mu \omega^{\mu}_{0} \, .
\end{equation}
\noindent This suggests that the space of solutions is a direct product of the flat and non-flat sectors. However, a more detailed analysis reveals that this is only true if the boundary conditions do not mix modes in the various sectors, see section \ref{sec:bc}.

Let us now find an expression for the symplectic flux that will allow us to determine the allowed boundary conditions. In order to do so, it is important to keep in mind that the modes $\delta B^{(\pm)}$ are constrained by the asymptotic equations of motion, and therefore obey \eqref{asymptotic constraint}. For example, in light-cone coordinates $(u,v)$ in which the boundary metric takes the form
\begin{equation}\label{bndy light cone}
    g^{(0)}_{ij} = \left(
                     \begin{array}{cc}
                       0 & 2 \\
                       2 & 0 \\
                     \end{array}
                   \right)
\end{equation}
\noindent these lead to
\beq
\delta B_{v}^{(+)} = \delta B_{u}^{(-)}=0\,.
\eeq
\noindent Taking the asymptotic constraints \eqref{asymptotic constraint} into account then, we find that the symplectic flux through $R$ is given by
\begin{align}\label{flux coeff}
\Phi
  ={}&
 \hat{\alpha}\int_{R}d^{2}x\,
	\varepsilon^{r\nu\lambda}\left( \delta_1 A^{(0)}_{\nu} \delta_2 A^{(0)}_{\lambda}-\delta_1 B_{\nu}^{(+)}\delta_2 B_{\lambda}^{(-)} + \delta_2 B_{\nu}^{(+)} \delta_1 B_{\lambda}^{(-)}  \right)
	\nonumber\\
	={}&
		\hat{\alpha}\int_{R}d^{2}x\,\varepsilon^{ij}\left( \delta_1 A^{(0)}_{i} \delta_2 A^{(0)}_{j}-\delta_1B_{i}^{(+)}\delta_2 B_{j}^{(-)} + \delta_2 B_{i}^{(+)} \delta_1 B_{j}^{(-)}  \right).
\end{align}

\subsection{Boundary conditions}\label{sec:bc}
As discussed above, demanding the vanishing of the symplectic flux gives us a useful way of classifying the boundary conditions. Momentarily giving up covariance in the boundary directions, in light-cone coordinates \eqref{bndy light cone} we find that possible local boundary conditions include
\begin{eqnarray}\label{bcs flat sector super}
A^{(0)}_{u} = W\bigl[A^{(0)}_{v}\bigr],\qquad  B_{u}^{(+)} = V\bigl[ B_{v}^{(-)}\bigr].
\end{eqnarray}

\noindent For general ``potentials" $W$ and $V$, such boundary conditions would correspond to multi-trace deformations in the dual CFT. For simplicity, let us focus on the linear case
\begin{eqnarray}\label{bcs flat sector}
\delta A^{(0)}_{u} = \bar{\beta}\, \delta A^{(0)}_{v},\qquad \delta B_{u}^{(+)} = \beta\, \delta B_{v}^{(-)},
\end{eqnarray}
\noindent for any constants $\beta,\bar\beta$. Note that $\beta=0,\infty$ correspond to chiral boundary conditions, while other values mix the modes and break covariance. We will refer to $\delta B_{u}^{(+)} = 0$ as Dirichlet and to $\delta B_{v}^{(-)} = 0$ as Neumann boundary conditions, in close analogy to the terminology commonly used for scalar fields in AdS. We will term the boundary condition $\delta B_{u}^{(+)} = \beta\, \delta B_{v}^{(-)}$ as ``mixed" when $\beta$ is finite. As usual, the boundary conditions with finite $\beta$ and $\bar\beta$ are related to double-trace deformations of the boundary theory \cite{Witten:2001ua,Berkooz:2002ug}, as we will review later on. Furthermore, we notice that, because $B_{u}^{(+)}$ and $B_{v}^{(-)}$ have scaling dimensions $\Delta_- = 1- \alpha$ and $\Delta_+ = 1 + \alpha$, respectively, the constant $\beta$ has dimension $\Delta_\beta = - 2\alpha$. The RG flow interpretation of double-trace deformations has been discussed in, for example, \cite{Witten:2001ua,Gubser:2002zh,Gubser:2002vv,Leigh:2003gk,Hartman:2006dy}.\footnote{This interpretation requires both end points of the RG flow to be well-defined, e.g. as in the case of scalar fields with masses close to Breitenlohner-Freedman bound in AdS. We shall see below that in the present case the Neumann theories are ill-defined so this picture does not strictly hold.} On the other hand, since $A^{(0)}_u$ and $A^{(0)}_v$ both have dimension one, the constant $\bar{\beta}$ is dimensionless.
Interestingly, we also note the possibility of a ``hybrid" boundary condition
\begin{equation}\label{gen hyb}
\delta A^{(0)}_{u} = \kappa\, \delta B_{u}^{(+)} \qquad \mbox{\textbf{and}}\qquad \delta A^{(0)}_{v} = \frac{1}{\kappa}\delta B_{v}^{(-)}\, ,
\end{equation}
\noindent that mixes the flat connections with the massive sector. Here, $\kappa$ is a constant of scaling dimension $\Delta_\kappa = \alpha$. Notice that, in view of the flatness condition on $\delta A^{(0)}$, \eqref{gen hyb} implies
\begin{equation}\label{hyb on b}
    \kappa^2\, \partial_v \delta B_u^{(+)} = \partial_u \delta B_v^{(-)}\, .
\end{equation}
\noindent In analogy with the linear boundary conditions discussed above, this hybrid boundary condition has the interpretation of a double-trace deformation. To our knowledge, the possibility of such boundary conditions has not been explicitly discussed in the literature.

It is now clear from the decomposition \eqref{omega split} and the analysis of the boundary conditions above that, as anticipated in \cite{Gukov:2004id}, the flat and massive sectors do not always decouple. In fact, for our hybrid boundary conditions \eqref{gen hyb} both sectors indeed interact with one another. The decoupling only occurs if one imposes boundary conditions which do not mix both sectors, i.e. if we impose  boundary conditions like those in \eqref{bcs flat sector}. This is because it is only in this case that the symplectic structure effectively splits as a direct sum of two independent pieces.

\subsection{One-point functions}
\label{sec:1pt}

As usual in the context of holography, the Maxwell-Chern Simons action \eqref{TME bulk action} must be supplemented by a series of boundary terms that serve two purposes: achieving a well-defined variational principle for a chosen set of boundary conditions, and removing divergences. We will refer to the latter as counterterms. We recall now that the first variation of the bulk action is given by \eqref{first variation action}. Evaluating this expression on-shell we find
\begin{equation}
\left.\delta I \right|_{os} = \frac{\hat{\alpha}}{2} \int_{\partial M}d^{2}x\,\sqrt{|\gamma|}\, \rho_{\mu}\epsilon^{\mu\rho\nu}\left(B_{\rho}- A^{(0)}_{\rho}\right)\delta A_{\nu}\, ,
\end{equation}

\noindent where the gauge field fluctuations are understood to be evaluated on the solution of the linearized equations of motion.\footnote{Since we are ignoring backreaction, the various metric quantities are always understood to be evaluated on their (fixed) background values.}  Employing the notation established above, we find
\begin{align}
\left.\delta I \right|_{os}
={}&
 -\frac{\hat{\alpha}}{2} \int_{\partial M}d^{2}x\,\varepsilon^{ij}\left(B_{i}- A^{(0)}_{i}\right)\left( \delta A^{(0)}_{j} + \delta B_{j}\right)
  \nonumber\\
 ={}&
  -\frac{\hat{\alpha}}{2} \int_{\partial M}d^{2}x\,\varepsilon^{ij}\left(B^{(+)}_{i}\delta B_{j}^{(-)} + B_{i}^{(-)}\delta B_{j}^{(+)}- A^{(0)}_{i}\delta A^{(0)}_{j}  \right)
  \nonumber\\
 &
  -\frac{\hat{\alpha}}{2}\lim_{r\to \infty} \int_{\partial M}d^{2}x\,\varepsilon^{ij} r^{\alpha}\left(B_{i}^{(+)}\delta A^{(0)}_{j} + A_{j}^{(0)}\delta B_{i}^{(+)}\right)\, ,
\end{align}
\noindent where in the last equality we used the restrictions placed by the asymptotic equations of motion on the $B^{(\pm)},\delta B^{(\pm)}$ modes (c.f. section \ref{subsection:asymptotic sols}). We note the presence (for any finite Maxwell coupling $q^2$) of the divergent term, which we cancel by the addition of a counterterm. Noticing that $\varepsilon^{ij} r^{\alpha}\left(B_{i}^{(+)}\delta A^{(0)}_{j} + A_{j}^{(0)}\delta B_{i}^{(+)}\right) = \delta \left(r^{\alpha}\varepsilon^{ij}A_{j}^{(0)}B_{i}^{(+)}\right)$ it is easy to check that the desired counterterm is given by the covariant expression
\begin{align}
I_{ct} = \frac{1}{2q^2} \int_{\partial M}d^{2}x\sqrt{|\gamma|}F^{i}A_{i}\, ,
\end{align}
\noindent where, as before, $\gamma$ is the determinant of the induced metric on the $r={\rm constant}$ surface, and we have defined
\begin{equation}
F^{i}\equiv \rho_{\mu}F^{\mu i}\, ,
\end{equation}
\noindent with $\rho_{\mu}$ the unit normal 1-form on the radial slices. Therefore, we have that
\begin{align}
\left.\delta\left( I + I_{ct}\right) \right|_{os}
 ={}&
  -\frac{\hat{\alpha}}{2} \int_{\partial M}d^{2}x\,\varepsilon^{ij}\left(B^{(+)}_{i}\delta B_{j}^{(-)} + B_{i}^{(-)}\delta B_{j}^{(+)}- A^{(0)}_{i}\delta A^{(0)}_{j}  \right)
 \end{align}
\noindent is finite as $r\to\infty$.
\subsubsection{Covariant boundary conditions}
In order to proceed further we need to discuss the additional \textit{finite} boundary terms needed in order to enforce different boundary conditions of interest. Confining ourselves to covariant terms for the moment, we consider the following quantities:
\begin{align}
B_{\pm}
 ={}&
  \mp \frac{1}{4\gc^{4}\hat{\alpha}}\int_{\partial M}d^{2}x\sqrt{|\gamma|}F^{i}\gamma_{ij}F^{j}\, ,
\\
B_{(0)} ={}&
  \frac{1}{2q^{2}}\int_{\partial M}d^{2}x\, \varepsilon^{ij}F_{i}A_{j}+\frac{1}{4}\int_{\partial M}d^{2}x\sqrt{|\gamma|}\gamma^{ij}\left(\frac{1}{q^4\hat\alpha}F_{i}F_{j}-\hat\alpha A_{i}A_{j}\right)\, .
\end{align}

\noindent Evaluating on-shell we find
\begin{align}\label{on shell covariant counter-terms}
\left. B_{\pm}\right|_{os}
 ={}&
 \pm  \frac{\hat\alpha}{2}\int_{\partial M}d^{2}x\sqrt{|g^{(0)}|}g^{(0)ij}B_{i}^{(-)}B_{j}^{(+)}
\nonumber\\
={}&
\pm  \frac{\hat\alpha}{2}\int_{\partial M}d^{2}x\,\varepsilon^{ij}B_{i}^{(+)}B_{j}^{(-)}
\\
 \left. B_{(0)} \right|_{os}
 ={}&
-\frac{\hat\alpha}{4}\int_{\partial M}d^{2}x\sqrt{|g^{(0)}|}g^{(0)ij}A_{i}^{(0)}A_{j}^{(0)}\, .
\end{align}
\noindent By taking linear combinations of these finite boundary terms we can achieve a variational principle well-suited for the various boundary conditions \eqref{bcs flat sector} of interest in the flat and gauge-invariant (massive) sectors. For example, we find
\begin{align}\label{variation with Bplus fixed}
\delta \left. \left(I + I_{ct}\pm B_{(0)} + B_{+}\right) \right|_{os}
={}&
     \hat{\alpha} \int_{\partial M}d^{2}x\sqrt{|g^{(0)}|}\left[\epsilon^{ij}B_{i}^{(-)}\delta B_{j}^{(+)} \mp A_{i}^{(0)}P_{\pm}^{ij}\delta A^{(0)}_{j}\right]\, ,
\end{align}
\noindent and
\begin{align}\label{variation with Bminus fixed}
\delta \left. \left(I + I_{ct}\pm B_{(0)} + B_{-}\right) \right|_{os}
={}&
     \hat{\alpha} \int_{\partial M}d^{2}x\sqrt{|g^{(0)}|}\left[\epsilon^{ij}B^{(+)}_{i}\delta B_{j}^{(-)}\mp A_{i}^{(0)}P_{\pm}^{ij}\delta A^{(0)}_{j}\right].
\end{align}
\noindent Now that we have identified the sources for the covariant boundary conditions, i.e. $\delta B^{(\pm)}_{i}$ and $\left(P_{\pm}\delta A^{(0)}\right)_{i} = g^{(0)}_{ij}P_{\pm}^{jk}\delta A^{(0)}_{k}$, we write the variation of the renormalized action $I_{ren}$ generically as
\begin{equation}\label{I ren gen}
\left.\delta I_{ren}\right|_{os} =  \int_{\partial M}d^{2}x\sqrt{|g^{(0)}|}\biggl[\langle \mathcal{O}^{(\pm)i}\rangle \delta B^{(\pm)}_{i} + \langle\mathcal{O}_{\pm}^{(0)\, i}\rangle \left(P_{\pm}\delta A^{(0)}\right)_{i}\biggr].
\end{equation}
\noindent Comparing with \eqref{variation with Bplus fixed} and \eqref{variation with Bminus fixed} and using the properties of $P_{\pm}$ we can read-off the one-point functions of the dual operators, and we obtain
\begin{align}\label{O b}
\langle \mathcal{O}^{(\pm)i}\rangle
 ={}&
  \pm \hat{\alpha}\,g^{(0)ij}B_{j}^{(\mp)}\, ,
  \\ \label{O A0}
 \langle\mathcal{O}_{\pm}^{(0)\, i}\rangle
 ={}&
 \mp \hat{\alpha} P_{\mp}^{ij}A_{j}^{(0)}\, .
\end{align}

Since $A_{i}^{(0)}$ is constrained by the flatness condition, the variational derivatives with respect to its components are ill-defined, and, as a consequence, the one-point functions \eqref{O A0} suffer from an ambiguity. However, this ambiguity is nothing but the one associated to the $U(1)$ gauge transformations. In other words, \eqref{O A0} are only defined up to the transformations $\delta A_i  = \partial_i \lambda$ that preserve the boundary conditions in the variational principle. See \cite{Marolf:2006nd} for a related discussion in the context of (pure) Maxwell fields.
\subsubsection{Symmetry-breaking boundary conditions}
Let us now turn to the less symmetric scenarios. First, we consider the case of ``mixed" boundary conditions, i.e. $B^{(+)}_u - \beta B^{(-)}_v = 0$, where $\beta$ is a finite dimensionful constant. It is clear that this requirement breaks both conformal and Poincar\'e symmetry, so we are allowed to write down the appropriate boundary terms simply in terms of the coefficients of the asymptotic expansion. As we shall see shortly, it is useful to generalize the above boundary condition and consider instead
\begin{equation}\label{mix bc J}
    B^{(+)}_u - \beta B^{(-)}_v = J_\beta\, ,
\end{equation}
\noindent where $J_\beta$ is an arbitrary fixed function of the boundary coordinates.  We ignore the contribution from the flat sector momentarily. Starting from the Neumann theory, i.e. the theory in which $B^{(-)}_i$ is fixed and whose action we denote by $I_N$, the boundary term we need to add in order to attain the mixed boundary condition is
\begin{equation}\label{I def mix}
    I_{def,\beta} = - \frac{\hat{\alpha}}{4 \beta} \int_{\partial M}d^{2}x\sqrt{|g^{(0)}|} \left(B^{(+)}_u \right)^2\, .
\end{equation}
\noindent In fact, using the variation of the Neumann action (\ref{variation with Bminus fixed}) and the explicit boundary term \eqref{I def mix}, we obtain
\begin{equation}\label{dI mixed}
    \delta\left(I_N + I_{def,\beta}\right) = \frac{\hat{\alpha}}{2 \beta} \int_{\partial M}d^{2}x\sqrt{|g^{(0)}|} B^{(+)}_u\left( \beta\, \delta B^{(-)}_v - \delta B^{(+)}_u \right),
\end{equation}
\noindent which is finite and stationary when the boundary condition \eqref{mix bc J} holds. Comparing \eqref{dI mixed} with \eqref{mix bc J}, we note that the quantity that is being held fixed in the variational principle is in fact $J_\beta$. This means that $J_\beta$ is to be interpreted as the source for the dual operator in the boundary theory. Given this, it follows from \eqref{dI mixed} that the one-point function in the presence of sources for the dual operator in the deformed theory is given by
\begin{equation}\label{1pt mix}
    \langle {\cal O}^{(-)}_u \rangle_{\beta} = - \frac{\hat{\alpha}}{2 \beta} B^{(+)}_u,
\end{equation}
\noindent where, as usual, $B^{(+)}_u$ must be thought of as a function of the source $J_\beta$ defined in \eqref{mix bc J}. Before constructing variational principles suitable for the remaining boundary conditions, we comment that the computation above provides a simple illustration of the well-known fact that linear boundary conditions of the form \eqref{mix bc J} correspond to double-trace deformations in the dual theory. The argument is as follows. First, we recall that in AdS/CFT the Neumann action $I_N$ is interpreted as the generating function for the operator associated to $B^{(+)}_u$ in the dual CFT. Then, the boundary term \eqref{I def mix} is transparently identified with a double-trace deformation for this operator. Moreover, the inclusion of \eqref{I def mix} implies that the original Neumann boundary condition needs to be shifted in such a way that the the modified action has an extremum. As noted above, the new boundary condition is nothing but the linear relation \eqref{mix bc J}, which completes the argument. It is worth commenting on the possibility of thinking of the \eqref{mix bc J} as a deformation of the Dirichlet theory. In such case, the boundary term that implements the shift in the boundary condition is quadratic in $B^{(-)}_v$, so it has dimension $2(1+\alpha)$. We see that the deformation is then irrelevant.

We now construct an appropriate action for the boundary condition
\begin{equation}\label{mix bc A0 J}
    A^{(0)}_u - \bar{\beta} A^{(0)}_v = J_{\bar \beta}
\end{equation}
\noindent where $\bar\beta$ is a non-zero dimensionless constant and $J_{\bar \beta}$ is a fixed arbitrary function of the boundary coordinates. In analogy with the previous case, $J_{\bar \beta}$ corresponds to the source of the dual operator. Note that since $\bar \beta$ is dimensionless the boundary condition \eqref{mix bc A0 J} for $J_{\bar \beta} = 0$ preserves scale invariance, yet it breaks Lorentz invariance. Once again, as a consequence of this, it is licit to write extra boundary terms which are not Lorentz densities. Moreover, because this boundary condition does not mix the flat and massive sectors, we concentrate on the flat connections and temporarily drop the contribution from the massive modes. Now, assuming that we start with an action $I^{(1)}_{ren}$ which attains an extremum when $P_- A^{(0)}$ is fixed, we find
\begin{equation}\label{dI1}
    \delta I^{(1)}_{ren} = \hat{\alpha} \int_{\partial M}d^{2}x\sqrt{|g^{(0)}|} \langle\mathcal{O}_{-}^{(0)\, i}\rangle (P_- \delta A^{(0)})_i = \frac{\hat{\alpha}}{2} \int_{\partial M}d^{2}x\sqrt{|g^{(0)}|} A^{(0)}_u \delta A^{(0)}_v
\end{equation}
\noindent as follows from \eqref{I ren gen} and \eqref{O A0}.  In this case, the boundary term that we need to add to $I^{(1)}_{ren}$ in order for \eqref{mix bc A0 J} to hold can be written as
\begin{equation}\label{I def bar beta}
    I_{def, \bar \beta} = - \frac{\hat{\alpha}}{4 \bar{\beta}} \int_{\partial M}d^{2}x\sqrt{|g^{(0)}|} \left( A^{(0)}_u \right)^2\, .
\end{equation}
\noindent In fact, with this choice the on-shell variation of the action reads
\begin{equation}\label{d I mix A0}
    \delta \left(I^{(1)}_{ren} + I_{def, \bar \beta}\right) = \frac{\hat{\alpha}}{2 \bar{\beta}} \int_{\partial M}d^{2}x\sqrt{|g^{(0)}|} A^{(0)}_u \left( \bar{\beta} \delta A^{(0)}_v - \delta A^{(0)}_u \right),
\end{equation}
\noindent as desired. As discussed above, the boundary condition \eqref{mix bc A0 J} is in one-to-one correspondence with the inclusion of the double-trace deformation \eqref{I def bar beta} in the dual theory. The relevant one-point function is given by
\begin{equation}\label{1pt mix A0}
    \langle {\cal O}^{(0)}_{+ u} \rangle_{\bar \beta} = - \frac{\hat{\alpha}}{2 \bar{\beta}} A^{(0)}_u\, .
\end{equation}

\noindent Once again, we mention that the one-point function \eqref{1pt mix A0} is only defined up to the appropriate $U(1)$ transformation.

Finally, we consider the ``hybrid" boundary conditions defined in \eqref{gen hyb}, which admit the obvious generalization
\begin{equation}\label{hyb bc J}
    A^{(0)}_u - \kappa B^{(+)}_u = J_\kappa\,, \qquad A^{(0)}_v - \kappa^{-1} B^{(-)}_v = \tilde{J}_\kappa \, ,
\end{equation}
\noindent where we take $J_\kappa$, $\tilde{J}_\kappa $ to be the sources of the dual operators. It is convenient to start with a renormalized action $I^{(2)}_{ren}$ such that
\begin{equation}\label{I ren for hyb}
    \delta I^{(2)}_{ren} = \frac{\hat{\alpha}}{2} \int_{\partial M}d^{2}x\sqrt{|g^{(0)}|} \left( A^{(0)}_u \delta A^{(0)}_v + B^{(-)}_v \delta B^{(+)}_u \right).
\end{equation}
\noindent With $I^{(2)}_{ren}$ as a starting point, the boundary term that implements hybrid boundary conditions is given by
\begin{equation}\label{I def hyb}
    I_{def, \kappa} = - \frac{\hat{\alpha}}{2 \kappa} \int_{\partial M}d^{2}x\sqrt{|g^{(0)}|} B^{(-)}_v A^{(0)}_u,
\end{equation}
\noindent as it follows from
\begin{align}\label{dI hyb}
    \delta\left(I^{(2)}_{ren} + I_{def, \kappa}\right)
    ={}&
     \frac{\hat{\alpha}}{2} \int_{\partial M}d^{2}x\sqrt{|g^{(0)}|} A^{(0)}_u \left( \delta A^{(0)}_v - \kappa^{-1} \delta B^{(-)}_v \right)
     \nonumber \\
    &-
     \frac{\hat{\alpha}}{2} \int_{\partial M}d^{2}x\sqrt{|g^{(0)}|} \kappa^{-1} B^{(-)}_v \left( \delta A^{(0)}_u - \kappa\, \delta B^{(+)}_u \right).
\end{align}

\noindent As pointed out before, the hybrid boundary conditions correspond to a double-trace deformation in the dual theory. Note that, in this case, the deformation \eqref{I def hyb} explicitly mixes the flat and massive sectors, so indeed these do not decouple in the theory defined by the hybrid boundary conditions. It follows from \eqref{dI hyb} that the one-point functions in the dual theory are given by
\begin{align}\label{O kappa 1}
    \langle {\cal O}^{(0)}_{+ u} \rangle_\kappa
    =
     \frac{\hat{\alpha}}{2} A^{(0)}_u \qquad \mbox{and}\qquad
    \langle {\cal O}^{(0)}_{- v} \rangle_\kappa
    =
     - \frac{\hat{\alpha}}{2 \kappa} B^{(-)}_v\, .
\end{align}
\noindent It is worthwhile noting that, since the $U(1)$ transformations do not preserve the boundary conditions \eqref{gen hyb}, the one-point functions \eqref{O kappa 1} are unambiguously defined.

Before closing this section, we emphasize that, provided $0 < \alpha < 1$, the boundary terms involved do not contain derivatives along the timelike direction. Given the results of \cite{Compere:2008us}, this strongly suggests that the bulk symplectic structure does not need to be supplemented by additional boundary contributions. This is indeed the case, as we will explicitly verify below. Specifically, we will check that, for $0 < \alpha < 1$, the bulk symplectic structure is finite and conserved for all the boundary conditions under scrutiny.

\subsection{The symplectic product}\label{sec: symp prod}
Recall that the symplectic structure is given by \eqref{omega split} with \eqref{omegas flat and non-flat}, i.e.
\begin{equation}\label{omega recap}
    \Omega = \hat{\alpha} \int_\Sigma d^2 x \sqrt{h} n_\mu \epsilon^{\mu\nu\rho}\delta_1 B_{\nu}\delta_2 B_{\rho} - \hat{\alpha}  \int_\Sigma d^2 x \sqrt{h} n_\mu \epsilon^{\mu\nu\rho}\delta_1 A^{(0)}_{\nu} \delta_2 A^{(0)}_{\rho}\, .
\end{equation}
\noindent As mentioned above, it turns out that the restriction $0 < \alpha < 1$ ensures that \eqref{omega recap} is finite for all the boundary conditions of interest, provided one imposes additional requirements on the solutions in the deep interior.

Quite generally, given a symplectic structure it is possible to endow the space of solutions with an inner product defined in terms of $\Omega$, as we now review briefly. A more detailed discussion can be found in \cite{Wald:1995yp}, for example. We start by complexifying the space of solutions and consider\footnote{The reader uneasy with the use of the complex conjugates in \eqref{ip} can think of using a basis of solutions in momentum space in which the modes are generically complex despite the fact that the field is real.}
\begin{equation}\label{ip}
    (A_1, A_2) = - i \Omega(A_1^*, A_2)\, .
\end{equation}
\noindent We will refer to \eqref{ip} as the symplectic product of the theory. One can verify that \eqref{ip} satisfies the expected properties of bi-linearity and Hermiticity, although in general it fails to be positive definite.

The inner product \eqref{ip} allows us to introduce some useful terminology. First, we shall term a given solution $A_0$ as \textit{normalizable} if $(A_0, A)$ is finite for all $A$. As stated above, in our particular setup this translates into a requirement on the fields in the deep interior. Second, we define a \textit{ghost} to be an excitation of definite positive(negative) frequency with negative(positive) norm. Here, we will use the definition of positive frequency associated to the timelike Killing vector of the relevant background geometry under consideration. For example, if $\partial_t$ is a vector field which is timelike everywhere, the solution $A$ is said to be a positive frequency solution if $\partial_t A = - i \omega A $ with $\omega > 0$. Third, we will refer to a solution $A_{gauge}$ as \textit{pure gauge} if $(A_{gauge}, A) = 0$ for all $A$.

It should be noted that the presence of ghosts in a given system is correlated with the lack of unitarity in the associated quantum theory. At the classical level, the presence of ghosts also signals pathologies since these give negative contributions to the energy.

\section{The dual field theory spectrum}\label{section:spectrum}
In this section we determine the spectrum of normalizable solutions of the MCS system in $AdS_3$, in both global and Poincar\'e coordinates, for the various boundary conditions of interest. As explained in section \ref{sec: symp prod}, by ``normalizable" we mean excitations that have finite symplectic product with all the modes. We mention that, while normalizability at the conformal boundary is guaranteed by restricting the coupling $\alpha$ defined in \eqref{rescaled CS coupling} to satisfy $0 < \alpha < 1$, normalizability at the interior is achieved by restricting the wave functions appropriately. More precisely, when the geometry is global AdS we shall require the wave functions to be smooth at the origin, as is customary. In the Poincar\'e AdS case, in addition to smoothness in the interior, we restrict the wave functions in such a way that no symplectic flux can leak through the Poincar\'e horizon.

As explained in section \ref{sec: prelim}, the connection splits into flat and ``massive" pieces, and we can solve the bulk equations of motion separately for each sector. Moreover, as discussed in section \ref{sec:bc}, these sectors decouple unless we impose the ``hybrid" boundary condition \eqref{gen hyb}. Our strategy to find the spectrum will be to focus on the massive and flat sectors separately, and incorporate the effects of the mixing only when we discuss the hybrid boundary conditions. For the sake of simplifying the exposition, we display the general solution to the equations of motion of the massive mode in appendix \ref{section:solutions}, while here we focus exclusively on imposing the appropriate boundary conditions.

\subsection{Global $AdS_{3}$} \label{sec: spec global adS}
We first consider the MCS theory in global $AdS_{3}$, whose line element is given by \eqref{Global AdS3}. Since the spacetime is topologically trivial, there is no room for holonomies and the connection must be smooth at the origin, where the vector field $\partial_x =(1/L)\partial \varphi$ becomes singular. As a result, in addition to normalizability we must impose $A[\partial_x] = A_x = 0$ at $\rho = 0$. It should be stressed that setting $A^{(0)}_\rho = 0$ \textit{everywhere} in the bulk is generically in conflict with smoothness. To see why, we note that this implies that the components of $A^{(0)}$ along the boundary directions are independent of $\rho$ everywhere, so any boundary condition other than $A^{(0)}_x |_{\partial M} = 0$ would yield singular configurations. Having said this, we initiate the study of the spectrum for all the boundary conditions of interest.
\subsubsection{Flat sector}
\label{spec flat sector global}

We first consider the flat sector. Since there are no holonomies, the flat connections can be written as
\begin{equation}\label{soln flat global}
    \delta A^{(0)}_\mu = \partial_\mu \lambda\, ,
\end{equation}
\noindent where $\lambda$ is smooth everywhere, with  $\partial_x \lambda = 0 $ at $\rho=0$. Recall that in our analysis of the symplectic flux, we encountered the allowed boundary condition  \eqref{bcs flat sector}, which in terms of the $(t,x)$ coordinates defined in appendix \ref{section:solutions} takes the form
\begin{equation}\label{bc flat t x}
   \left.\left( \delta A^{(0)}_t - \hat{\beta}\, \delta A^{(0)}_x \right)\right|_{\partial M} = 0\, ,
\end{equation}
\noindent where $\hat{\beta} = (\bar{\beta}-1)/(\bar{\beta} +1)$ is a (possibly vanishing or infinite) constant. Fourier-decomposing $\lambda$ as $\lambda = e^{-i \omega t + i k x} \hat \lambda(k,\omega)$ with $k\in\mathds{Z}$, and using \eqref{soln flat global} in \eqref{bc flat t x}, we learn that the frequencies must satisfy
\begin{equation}\label{freq global}
    \omega = - \hat{\beta} k\, ,
\end{equation}
\noindent which determines the spectrum of the flat sector. As is well-known \cite{Witten:1988hf, Moore:1989yh,Elitzur:1989nr,Wen:1991mw,Balachandran:1991dw,Bos:1989kn,Bos:1989wa,Schwarz:1979ae}, the degrees of freedom of the flat sector reside exclusively on the spacetime boundary,\footnote{In particular, if $\lambda$ goes to zero at the boundary the flat connections are pure gauge.} a fact that we will briefly review in appendix \ref{section:sym bndy}.   We will consider the flat solutions for hybrid boundary conditions in the next subsection.

\subsubsection{Massive sector}
Focusing now on the massive sector we use the ansatz \eqref{ansatz global AdS}, in which case the solution is given by \eqref{solution b in global AdS3}-\eqref{the cs 2} in terms of functions $F(\omega,\pm k,\pm\alpha;\rho)$. We observe that only the $F(\omega,|k|,\alpha ;\rho)$ profiles are regular in the interior ($\rho \to 0$). Hence, for $k<0$, we take the $F(\omega,-k,\alpha ;\rho)$ solution. Consequently, we will write the general solution which is smooth in the interior of $AdS_3$ as
\begin{align}\label{regular solution b in global AdS3}
b_{u} ={}&
C_{u}F( \omega, |k|,\alpha;  \rho)
\\
b_{v} ={}&
C_{v}F(  \omega, |k|, -\alpha;  \rho)\, ,
\end{align}
\noindent where $(u,v)$ are the light-cone coordinates defined in \eqref{nc}, $F( \omega, k,\alpha;  \rho)$ is defined as in \eqref{F hyp}, and
\begin{equation}\label{relation coeff global AdS}
   \frac{C_{v}}{C_{u}} = \frac{k+\omega- s(k) \alpha}{k-\omega+ s(k) \alpha}\, .
\end{equation}

\noindent Here, $s$ denotes the sign function, i.e. $s(k) = 1 $ for $k \geq 0$ and $s(k) = - 1$ for $k < 0$. The $b_{\rho}$ component is obtained from $b_{u}$ and $b_{v}$ via \eqref{radial constraint global ads} and it is subleading with respect to them near the conformal boundary of $AdS_{3}$. Expanding $F(\omega,k,\alpha ;\rho)$ near $\rho = \infty$ and using \eqref{regular solution b in global AdS3}-\eqref{relation coeff global AdS}, we learn that the relevant coefficients in the asymptotic expansion are
\begin{align}\label{explicit coeffs b+}
    b^{(+)}_u
     ={}&
      C_u C(\alpha, |k|, \omega)\,,
      &    b^{(+)}_v ={}&0 \, ,
\\
\label{explicit coeffs b-}
    b^{(-)}_v
    ={}&
      C_u \frac{k+\omega- s(k) \alpha}{k-\omega+ s(k) \alpha} C(-\alpha, |k|, \omega)\,,
     &   b^{(-)}_u ={}&0\, ,
\end{align}
\noindent where
\begin{equation}\label{C alpha}
    C(\alpha, k, \omega) = \frac{\Gamma(k+1) \Gamma(1+\alpha)}{\Gamma\left(\displaystyle{1+\frac{ k + \alpha - \omega}{2}}\right) \Gamma\left(\displaystyle{1+\frac{ k + \alpha + \omega}{2}}\right)}\, .
\end{equation}

A choice of asymptotic boundary conditions will constrain the allowed values of $(\omega,k)$, corresponding to the normal modes of the system. In the Dirichlet case $(\beta=0)$ the source is identified with $b^{(+)}_{u}$ and the normal modes are given by the zeros of $C(\alpha, |k|, \omega)$, located at
\begin{equation}\label{omega D2}
    \omega^{\pm}_{nk} = \pm \left(2 n + |k|  +\alpha \right)\, ,\qquad   n = 1,2,\ldots
\end{equation}
\noindent and the zeros of the denominator in \eqref{relation coeff global AdS}, located at
\begin{alignat}{2}\label{omega D zero modes}
    \omega^{+}_{0k}
    ={}&
   +(  |k| + \alpha)    &\phantom{aaaa}&{\rm for} \ k > 0
     \\
     \omega^{-}_{0k}
     ={}&
      -(|k| + \alpha)  &\phantom{aaaa}&{\rm for} \ k < 0\, .
      \label{omega D zero modes 2}
\end{alignat}
\noindent We notice that the normal modes with $n \geq 1$ are doubly degenerate, with each frequency attained for both $k$ and $-k$, while $\omega^\pm_{0k}$ occur only once.

\begin{figure}[htb]
\center
\subfigure[][]{
\label{w mix bc}
\includegraphics[width=0.4\linewidth]{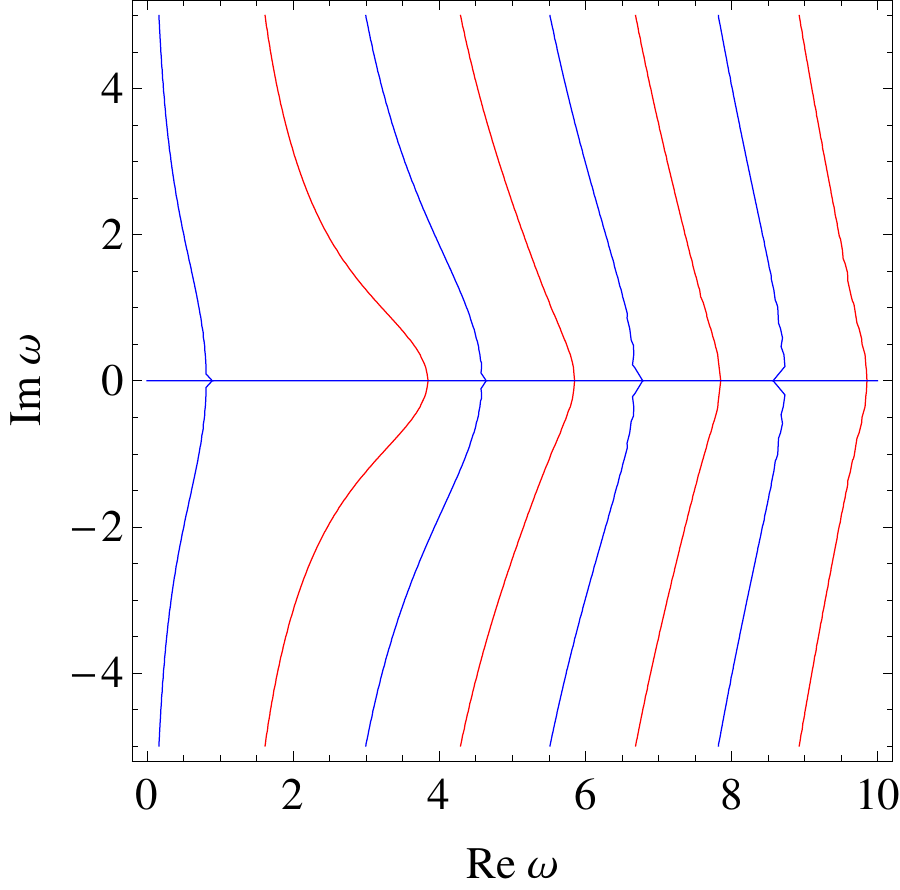}
}\qquad\qquad
\subfigure[][]{
\label{w mix bc k neg}
\includegraphics[width=0.4\linewidth]{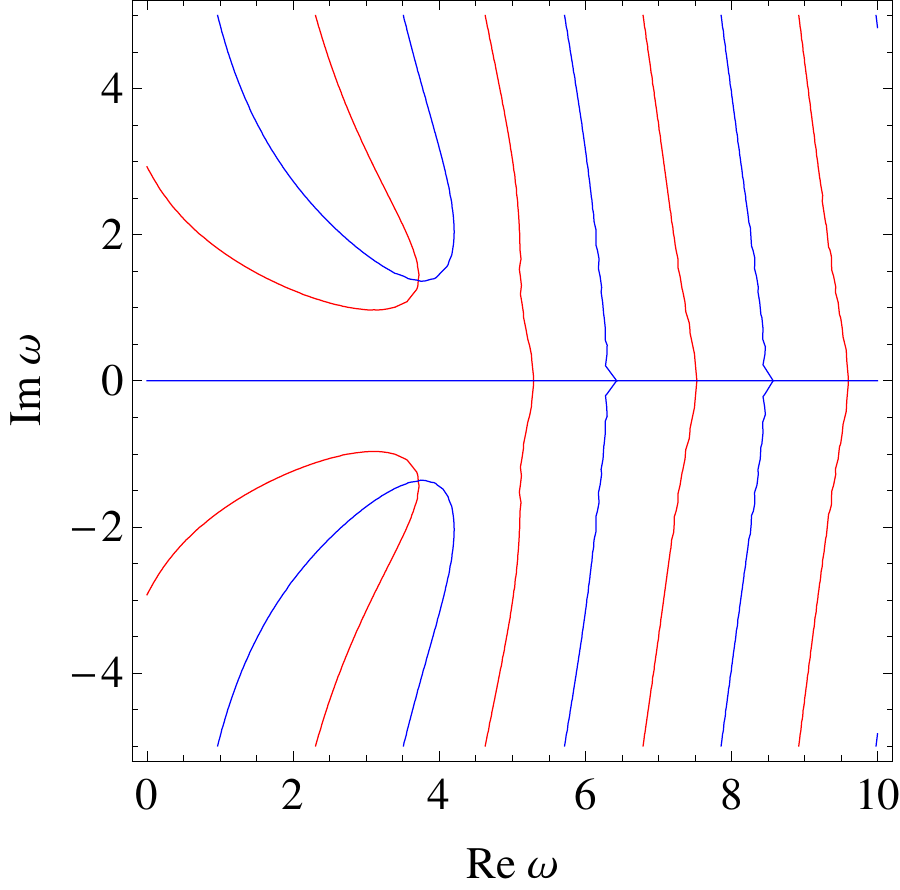}}
\caption{\ref{w mix bc}:  We plot in red/blue the solution of the real/imaginary part of equation \eqref{bc eq for w} in the complex-$\omega$ plane for $\{k =2,\,\alpha = 0.2,\,\beta = 1.7\}$. We observe that these solutions only intersect for $\mbox{Im}(\omega) = 0$, which illustrates the fact  that \eqref{bc eq for w} has only real solutions for $\beta > 0$. \ref{w mix bc k neg}: For $\{k =-2,\,\alpha = 0.2,\, \beta = -1.7\}$, we plot in red/blue the solution of the real/imaginary part of equation \eqref{bc eq for w} in the complex-$\omega$ plane. We note that in this case there are complex frequency solutions. }\end{figure}

\noindent Similarly, for Neumann boundary condition $(b^{(-)}_v = 0)$ we find the eigenfrequencies
\begin{equation}\label{omega N2}
    \omega^{\pm}_{nk} = \pm \left(2 n + |k|  - \alpha \right)\, ,\qquad   n = 1,2,\ldots
\end{equation}
\noindent in addition to
\begin{alignat}{2}
    \omega^{-}_{0k}
    ={}& \alpha - |k|  &\phantom{aaaa}&{\rm for} \ k > 0
     \\
     \label{omega N zero modes k <0}
     \omega^{+}_{0k}
     ={}&
      |k| - \alpha &\phantom{aaaa}&{\rm for} \ k < 0\, .
\end{alignat}
\noindent More generally, the boundary condition $b^{(+)}_u = \beta\, b^{(-)}_v $ for finite $\beta$ gives
\begin{equation}\label{bc eq for w}
    C(\alpha, |k|, \omega) - \beta \frac{k+\omega- s(k) \alpha}{k-\omega+ s(k) \alpha} C(-\alpha, |k|, \omega) = 0\, .
\end{equation}
\noindent For generic $\beta$, we will proceed numerically, examining the structure of the solutions of \eqref{bc eq for w} in the complex-$\omega$ plane as a function of $k$, $\beta$ and $\alpha$. For $\beta > 0$ and all values of $k$, we find an infinite discrete set of real frequency solutions, in analogy to the Dirichlet and Neumann cases. Now, while for $\beta < 0$ and $k>0$ all frequencies are real, for $\beta < 0$ and $k<0$ a pair of complex solutions occurs in addition to the series of real solutions. Notice that, with the exception of $\omega$, all the parameters in \eqref{bc eq for w} are real, which implies that complex solutions must appear in complex conjugate pairs. These complex solutions go off to $\pm i \infty$ as $\beta \rightarrow 0$, in agreement with our analysis for Dirichlet boundary conditions.  See figures \ref{w mix bc} and \ref{w mix bc k neg}. The complex frequency solutions signal an instability of the system, since some perturbations can grow exponentially with time. This instability is associated with ghosts, as we will see in section \ref{ip massive global}. We stress that, aside from the existence of complex frequencies, there is nothing particularly special about $\beta < 0$. In fact, we will see below that all values of $\beta \neq 0$ are qualitatively equivalent, since they all yield ghosts.

\begin{figure}[htb]
\center
\subfigure[][]{
\label{tach hyb real}
\includegraphics[width=0.4\linewidth]{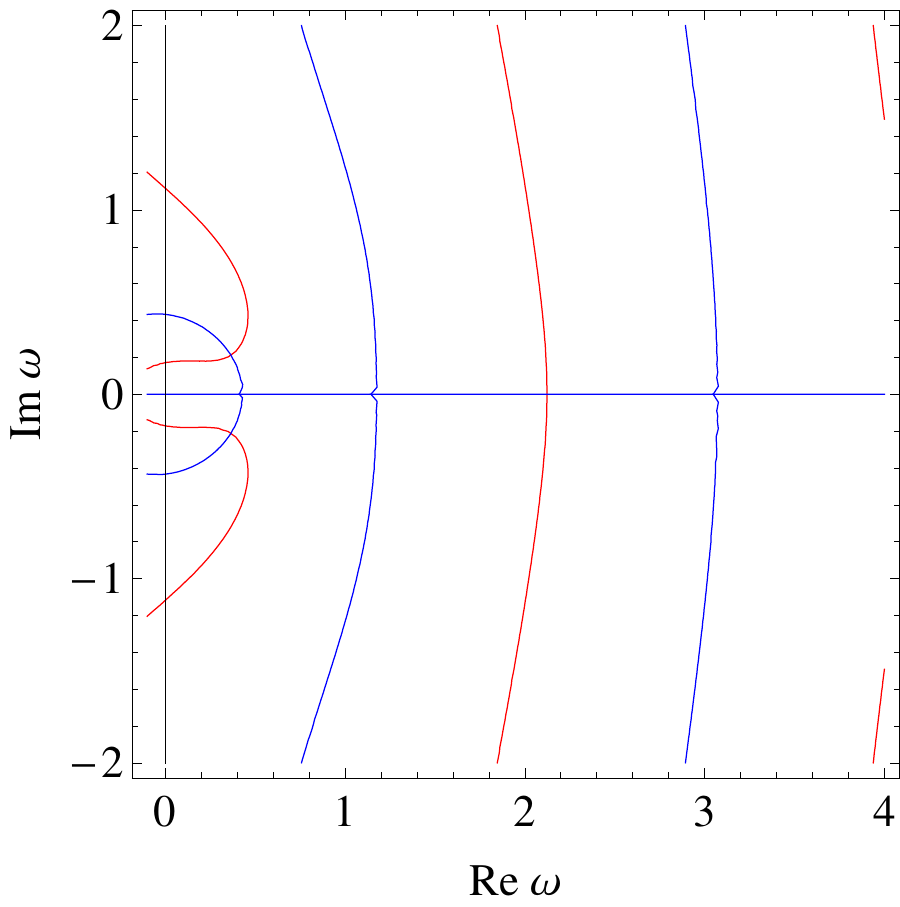}
}\qquad\qquad
\subfigure[][]{
\label{tach hyb cplex}
\includegraphics[width=0.4\linewidth]{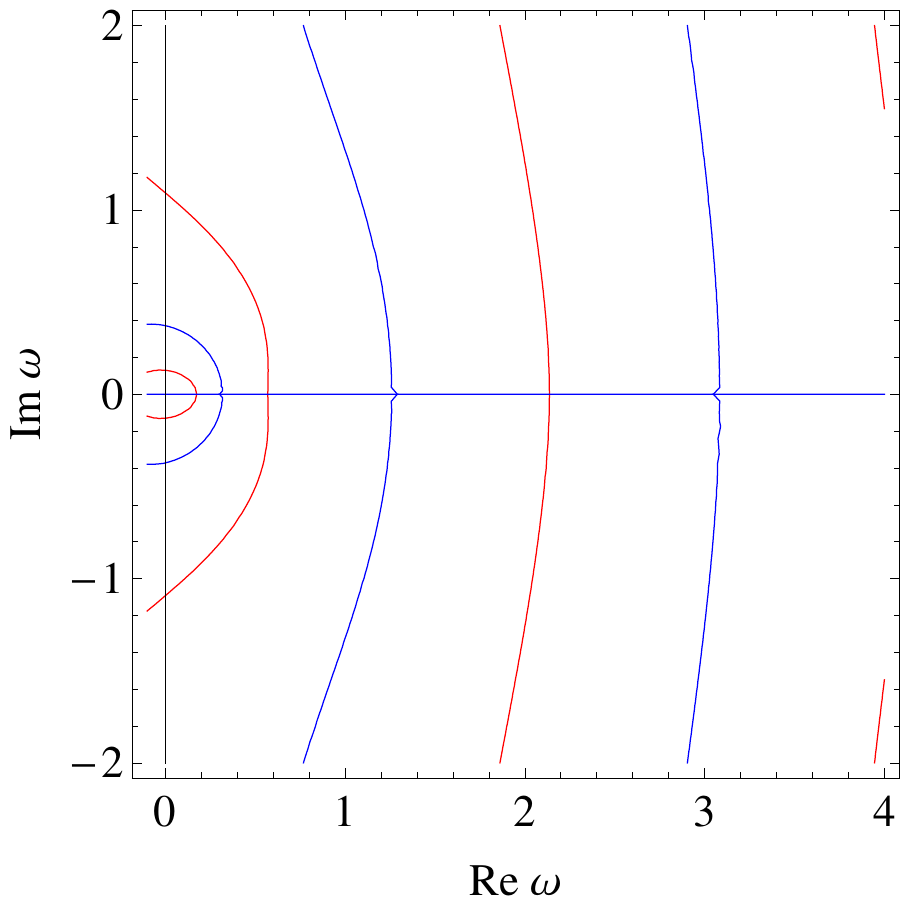}}
\caption{\ref{tach hyb real}: We plot in red/blue the solution of the real/imaginary part of equation \eqref{hyb bc coeff} in the complex-$\omega$ plane for $\{k =1,\,\alpha = 0.8,\,\kappa = 1.0\}$. We note that there are real solutions but also a pair of complex solutions near $|\omega| = 0$.
\ref{tach hyb cplex}: Solutions for $\{k =1,\,\alpha = 0.8,\, \kappa = 0.9\}$. We observe that the complex solutions become real, which shows that for $\alpha = 0.8$ the critical value of $\kappa$ is near $|\kappa_c| = 0.95\,$.}
\end{figure}

Finally, we consider the hybrid boundary conditions \eqref{gen hyb}. As noted in section \ref{sec:bc}, the condition \eqref{gen hyb} along with the flatness of $\delta A^{(0)}$ imply the extra requirement \eqref{hyb on b}, which in view of our mode decomposition translates into
\begin{equation}\label{hyb bc coeff}
    \kappa^2 \frac{k + \omega}{k -\omega} C(\alpha, |k|, \omega) - \frac{k+\omega- s(k) \alpha}{k-\omega+ s(k) \alpha} C(-\alpha, |k|, \omega) = 0\, .
\end{equation}
Thus, the spectrum of frequencies is given by the solutions of \eqref{hyb bc coeff} provided the flat components of the connection are related to the massive ones by \eqref{gen hyb}. Lacking an analytic solution of \eqref{hyb bc coeff} for finite $\kappa$, we proceed numerically. Studying \eqref{hyb bc coeff} for various values of the parameters, we find that generically there is an infinite set of real solutions. Additionally, a pair of complex solutions occurs when $k > 0$ and $|\kappa| > |\kappa_c|$, where $\kappa_c$ is an increasing function of $\alpha$ and $k$. See figures \ref{tach hyb real}, \ref{tach hyb cplex} for an illustration of this fact. We have also verified numerically that the complex solutions go off to $\pm i \infty$ as $|\kappa|$ approaches infinity, consistent with the Dirichlet result. As in the case of mixed boundary conditions, the complex frequency solutions correspond to a dynamical instability of the system that is associated to ghosts. We shall also find that the all finite values of $\kappa$ yield ghosts, in agreement with the CFT unitarity bound.

As mentioned above, given a solution of \eqref{hyb bc coeff} the components of the flat connection are uniquely determined by \eqref{gen hyb}.
It is worth mentioning that with these boundary conditions the chiral currents acquire a non-vanishing expectation value. See section \ref{sec:1pt}.

\subsection{Poincar\'e patch of  $AdS_{3}$}\label{sec: spec PP}
We now carry out the study of the spectrum of normalizable excitations for the boundary conditions of interest in the Poincar\'e patch of $AdS_{3}$. As in the global AdS case, normalizability at the conformal boundary is guaranteed by the restriction $0 < \alpha < 1$. On the other hand, the treatment of the Poincar\'e horizon turns out to be more delicate as we will discuss in detail below.
\subsubsection{Flat sector}
Let us first consider the flat sector. As mentioned in section \ref{subsection:symplectic flux}, when the geometry is the Poincar\'e patch of $AdS_3$, symplectic flux can generically leak through the Poincar\'e horizon. In the flat sector, the easiest way to see this is to note that in this sector the theory is actually topological, so there is no difference between the Poincar\'e horizon and the conformal boundary. From our experience with the latter, we conclude that good boundary conditions in the flat sector correspond to fixing half of the connection on the Poincar\'e horizon. We will impose the condition
\begin{equation}\label{bc A0 at P hor}
    \left. \delta A^{(0)}_x \right|_{z = \infty} = 0\, .
\end{equation}
As reviewed in appendix \ref{section:sym bndy}, when fixing the spatial part of $A^{(0)}_i$, the degrees of freedom that reside at the Poincar\'e horizon become pure gauge, which allows us to focus on the physics at the boundary. Note however that \eqref{bc A0 at P hor} can be generalized in the same way as the boundary conditions discussed in section \ref{sec:bc}. Also, in analogy with the global case, we see that $U(1)$ transformations that set $A_z = 0$ everywhere in the bulk generically do not preserve the boundary condition \eqref{bc A0 at P hor}, so they are not allowed symmetries of the system.

From the above discussion, it is clear that the spectrum of the flat connections in the Poincar\'e case is analogous to the one in global AdS discussed in section \ref{sec: spec global adS}. In particular, the frequencies are fixed as \eqref{freq global} as a consequence of the boundary conditions at the conformal boundary, which are identical to the ones we consider in the Poincar\'e patch. Note however that in the present case the spatial momentum $k$ is not quantized, so the spectrum of eigenfrequencies is continuous.
\subsubsection{Massive sector}
Let us now focus on the massive sector. In order to solve the equations of motion, we use the mode decomposition $\delta B_\mu =  e^{i( k_u u + k_v v )} b_\mu\, $; see appendix \ref{sec: soln PP} for the explicit solutions. We classify the modes according to the value of $m^2:= -k_u k_v = \omega^2 - k^2$ as: timelike ($m^2 > 0$), lightlike ($m^2 = 0$), and spacelike $(m^2 < 0)$.

From \eqref{gen asympt} it follows that the asymptotic expansion of the solution for the massive mode reads (after noting that near the boundary we have $r = 1/z$)
\begin{equation}\label{asympt PP}
    b_\mu = z^{-\alpha} b^{(+)}_\mu + z^{\alpha} b^{(-)}_\mu + \mathcal{O}\left(z^{1-\alpha}\right).
\end{equation}
Here $z$ is the radial variable defined in \eqref{ds2}. Note that under the isometry \eqref{dilation PP}, the coefficients in \eqref{asympt PP} scale as
\begin{equation}\label{scaling}
    b^{(+)}_\mu \rightarrow c^{\alpha-1} b^{(+)}_\mu~,  \qquad b^{(-)}_\mu \rightarrow c^{-\alpha-1} b^{(-)}_\mu~,
\end{equation}
\noindent in agreement with our discussion of section \ref{subsection:conf dims} regarding the conformal dimensions of the dual operators.

Having said this, let us consider the spectrum of timelike modes, whose radial profile is given by \eqref{soln tl PP}. Comparing \eqref{soln tl PP} with \eqref{asympt PP}, we read-off
\begin{align}\label{}
    b^{(+)}_{u}
     ={}&
      k_u C(\vec{k}) \frac{2^{1+\alpha} m^{-(\alpha+1)}}{\Gamma(-\alpha)}\,,&
       b^{(+)}_v ={}&
        b^{(+)}_z = 0\, ,
\\
    b^{(-)}_v
    ={}&
     k_v A (\vec{k}) \frac{2^{1-\alpha} m^{\alpha-1}}{\Gamma(\alpha)}\,,&
      b^{(-)}_u ={}&
       b^{(-)}_z = 0\, .
\end{align}
\noindent Thus, $C(\vec{k}) = 0$ corresponds to Dirichlet and $A(\vec{k})= 0$ to Neumann boundary conditions. We also find that mixed boundary conditions imply
\begin{equation}\label{mix bc}
    C(\vec{k}) = \beta \frac{k_v}{k_u} \frac{\Gamma(-\alpha)}{4^\alpha \Gamma(\alpha)} m^{2\alpha} A(\vec{k})\,,
\end{equation}
\noindent while hybrid boundary condition translate into
\begin{equation}\label{hybrid bc}
    C(\vec{k}) = \kappa^{-2} \frac{\Gamma(-\alpha)}{4^\alpha \Gamma(\alpha)} m^{2\alpha} A(\vec{k})\, .
\end{equation}

We stress that the timelike modes above oscillate rapidly near $z= \infty$. As a result, one can construct wave packets that behave smoothly near the Poincar\'e horizon. Alternatively, one can work with the modes as they stand and treat their symplectic products in the appropriate distributional sense, and this is the strategy we adopt below. More precisely, in section \ref{sec ip PP} we find that the timelike modes are in fact (plane wave-)normalizable for all the boundary conditions of interest.

\begin{figure}[htb]
\center
\subfigure[][]{
\label{w real tach}
\includegraphics[width=0.42\linewidth]{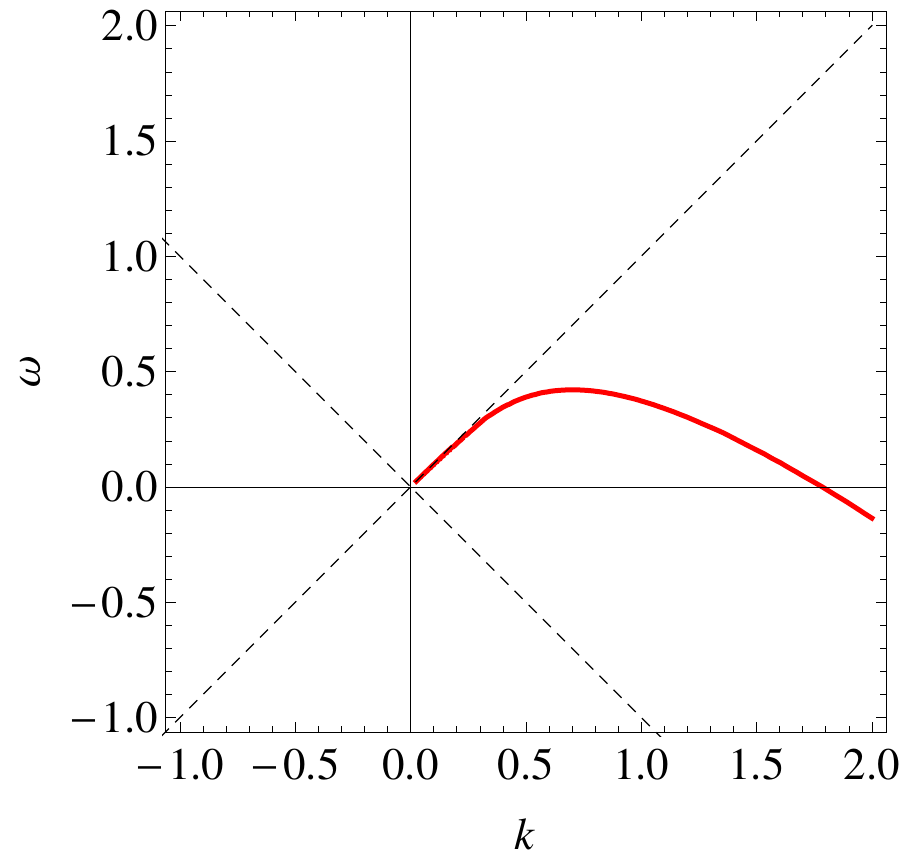}
} \qquad \qquad
\subfigure[][]{
\label{w cplex tach}
\includegraphics[width=0.4\linewidth]{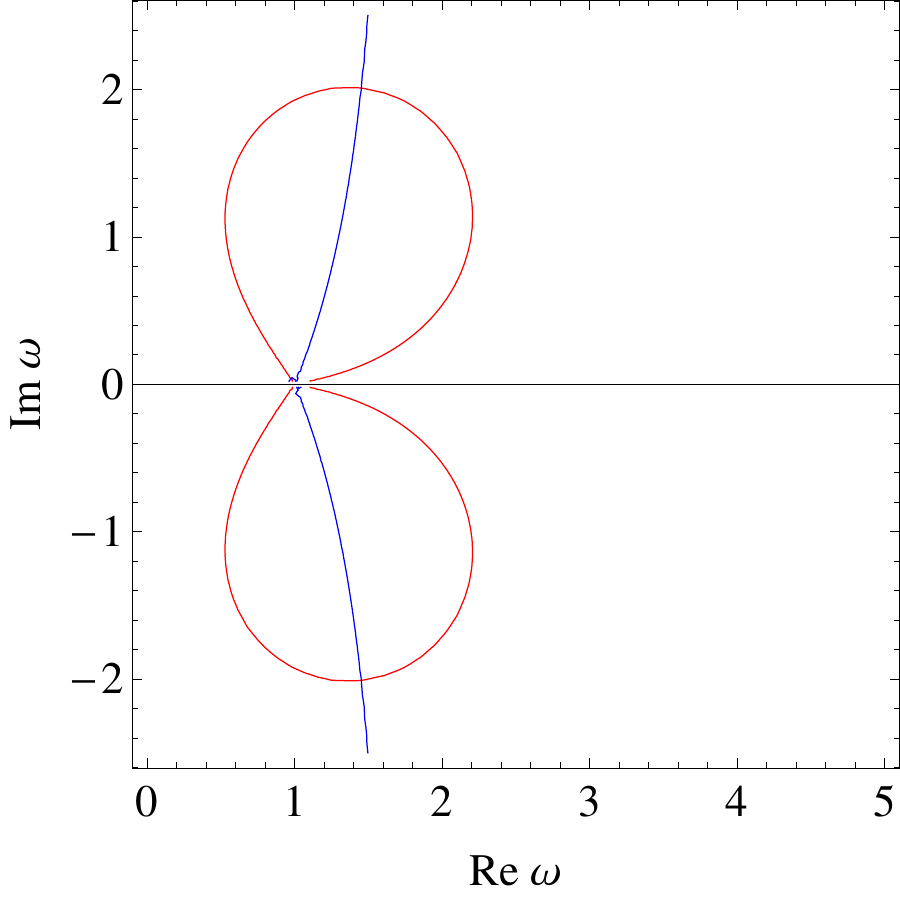}}
\caption{\ref{w real tach}: We plot in red the real solutions of \eqref{mix bc tach PP} in the $(k,\omega)$ plane for $\{\tilde{\beta} = 0.5\,,\alpha = 0.6\}$. The dashed line corresponds to the light-cone in momentum space. \ref{w cplex tach}: For $\{\tilde{\beta} = -0.5,\,\alpha = 0.6,\, k = -1\}$, we plot in red/blue the solutions to the real/imaginary part of \eqref{mix bc tach PP} in the complex-$\omega$ plane. Complex solutions are given by the intersection of both lines at $\omega \approx 1.45 \pm i 2.01\,$. This implies $p \approx 2.18 - i 1.34$ so $\mbox{Re}(p) > 0$, consistent with the assumption under which the solution is regular at the Poincar\'e horizon. }
\end{figure}

We now study the existence of spacelike solutions, whose profiles are given by \eqref{soln sl PP}. Taking $\mbox{Re}( p) >0$ by convention, we see that unless we set $C(\vec{k}) = 0$ in \eqref{soln sl PP}, the solutions blow up exponentially at the horizon ($z=\infty$) and are thus non-normalizable. Therefore, we set  $C(\vec{k}) = 0$ which implies that the coefficients of the asymptotic expansion for the spacelike solution can be written as
\begin{align}\label{bplus tach}
    b^{(+)}_u
     ={}&
       A(\vec{k}) k_u 2^{\alpha} p^{-\alpha-1} \Gamma(1+\alpha)\,,& b^{(+)}_v ={}& b^{(+)}_z = 0\, ,
    \\
\label{bminus tach}
    b^{(-)}_v ={}&
     A(\vec{k}) k_v  2^{-\alpha} p^{\alpha-1} \Gamma(1-\alpha)\,,& b^{(-)}_u ={}& b^{(-)}_z = 0\, .
\end{align}
\noindent Both Dirichlet and Neumann boundary conditions require $A(\vec{k}) = 0$, so in these cases spacelike solutions do not exist. Mixed boundary conditions $b^{(+)}_u = \beta b^{(-)}_v$, in turn, imply the relation
\begin{equation}\label{mix bc tach PP}
    \tilde{\beta} = \frac{(k-\omega)^{1-\alpha}}{(k+\omega)^{1+\alpha}}\, ,
\end{equation}
\noindent where we have defined $\tilde{\beta} = 4^{-\alpha} \frac{\Gamma(1-\alpha)}{\Gamma(1+\alpha)} \beta$. Spacelike solutions are then in one-to-one correspondence with the solutions of \eqref{mix bc tach PP}, which we now study. First, we observe that regularity at transverse infinity, $|x| \rightarrow \infty$, requires $k \in \mathbb{R}$. On the other hand, recall that we derived \eqref{mix bc tach PP} only under the assumption $\mbox{Re}(p) >0$, so in principle complex frequency solutions are allowed and their existence is exclusively dictated by \eqref{mix bc tach PP}. Examining \eqref{mix bc tach PP} it is not hard to conclude that for all $\beta >0$ there are real solutions in the region $k-\omega>0$, $k+\omega >0$; see figure \ref{w real tach}. On the other hand, if $\beta < 0$ real solutions are ruled out, but we find instead a pair of complex-frequency solutions that are conjugate to each other, see figure \ref{w cplex tach}. The fact that our results depend on the sign of $\beta$ only can be easily understood in terms of the scaling symmetry \eqref{dilation PP}, which acts non-trivially on $\beta$.

Similarly, for spacelike solutions satisfying the ``hybrid'' boundary condition \eqref{hyb on b} we have
\begin{equation}\label{hyb bc tach PP}
    \tilde{\kappa}^2 = (k^2 - \omega^2)^\alpha\, ,
\end{equation}
\noindent where we defined $\tilde{\kappa}^2 = 4^\alpha \frac{\Gamma(1+\alpha)}{\Gamma(1-\alpha)} \kappa^2 $. Since $\tilde{\kappa}^2 > 0$, it follows that $ \omega^2 = k^2 - \tilde{\kappa}^{2/\alpha}\,$. Now, because $k$ can be arbitrarily small, we find real as well as imaginary frequency solutions for all values of $\kappa$. In analogy to the mixed boundary conditions studied above, we can use the scaling symmetry \eqref{dilation PP} to set $\kappa$ to any desired value. Furthermore, in this case the spectrum is insensitive to the sign of $\kappa$ due to the structure of the boundary condition \eqref{hyb on b}.

Finally, we discuss the lightlike modes. For the right-moving modes, i.e. those with $k_{v}=0$, the general solution is given in \eqref{soln PP kv=0}. Examining the expression for the inner product, we conclude that the norm of the right-moving modes diverges if $b_z \neq 0$. Therefore, we find that right-moving modes are only allowed for Neumann boundary conditions. In this case, they read
\begin{equation}\label{RM Neumann}
    \delta B = A(k_u) z^{-\alpha} e^{i k_u u} du\, .
\end{equation}

\noindent We emphasize that the solution \eqref{RM Neumann} is smooth at the Poincar\'e horizon. Similarly, the left-moving modes \eqref{soln PP ku=0} are only normalizable only for Dirichlet boundary conditions, in which case they can be written as
\begin{equation}\label{LM Dirichlet}
    \delta B = C(k_v) z^{\alpha} e^{i k_v v} dv\,.
\end{equation}
\noindent Note however that in this case they fail to be smooth at $z = \infty$, which removes them from the spectrum.

\section{Evaluating the symplectic product}\label{section:norms}
Next, we compute the symplectic product of the various solutions found in section \ref{section:spectrum}. The emphasis will be on determining the existence of ghosts, which, as stated above, correspond to positive (resp. negative) frequency modes having negative (resp. positive) norm. According to CFT considerations regarding unitarity bounds for vector operators, we expect the theories in which $B^{(+)}$ fluctuates to contain ghosts. Up to certain subtleties present in the Poincar\'e patch, we will verify that the expected ghosts arise in the bulk, consistent with the field theory result. In addition, we will also find ghosts in the flat sector for a certain class of double-trace boundary conditions; the latter are not related to unitarity bounds of the kind mentioned above. The presence of these ghosts should not be at all surprising, however, since the symplectic structure restricted to the flat sector is not manifestly positive definite, see e.g. \eqref{omega split}. We find it convenient to study first the flat sector separately, assuming that we have chosen boundary conditions which decouple this sector from the massive one. The results for the massive sector and the mixed hybrid case will be presented later in this section.

\subsection{Flat sector}\label{ip flat sector}
We start by discussing the case of global AdS. The symplectic product is evaluated on a slice of constant $t$, so we have $\sqrt{h} n_\mu = \sqrt{g} \delta_\mu^t$. Then, using \eqref{omega recap} and \eqref{ip} the symplectic product reads
\begin{equation}\label{ip flat 1}
    (A_1,A_2) = - i \hat{\alpha} \int d^2x\, \varepsilon^{t \lambda \nu} \left(\delta_1 A^{(0)}_{\lambda}\right)^* \delta_2 A^{(0)}_{\nu}\,.
\end{equation}
\noindent Using the solution \eqref{soln flat global} and the mode decomposition $\lambda = e^{-i \omega t + i k x} \hat \lambda$, it is straightforward to arrive at
\begin{equation}\label{ip flat 2}
    (A_1,A_2) = - 2 \pi \hat{\alpha}\, \delta_{k_1, k_2} k_1 e^{i t (\omega_1 - \omega_2)} \int_0^\infty d\rho \left(\hat{\lambda}_2\partial_\rho \hat{\lambda}_1^*  -  \hat{\lambda}_1^* \partial_\rho \hat{\lambda}_2\right).
\end{equation}
\noindent  
Upon using \eqref{freq global} in \eqref{ip flat 2} we see that the time dependence in the symplectic product cancels out, as required by conservation of the symplectic structure. Finally, integrating by parts the first term in \eqref{ip flat 2} and using the smoothness condition $\delta A_x^{(0)} = 0 $ at $\rho = 0\,$, we conclude
\begin{equation}\label{ip flat 3}
    (A_1,A_2) = 2 \pi \hat{\alpha}\, \delta_{k_1, k_2} \frac{\omega_1}{\hat{\beta}} \bigl| \hat{\lambda}_{1, \partial}  \bigr|^2\, .
\end{equation}
\noindent where $\hat{\lambda}_{\partial} = \hat{\lambda} \bigr|_ {\partial M}$ are the (finite) boundary values of the Fourier components of $\lambda$. Note that \eqref{ip flat 3} is manifestly finite and conserved, as promised.
We observe that the symplectic product is local on the boundary values of $\lambda$, as expected in a topological theory with a boundary. In other words, flat connections for which $\lambda$ vanishes on the boundary are pure gauge degrees of freedom. Moreover, for the boundary condition $\bigl.\delta A^{(0)}_x \bigr|_{\partial M} = 0$, i.e. $k=0$, we also find that the flat sector becomes pure gauge. We refer the reader to appendix \ref{section:sym bndy} for a discussion on gauge symmetries.
Recall that ghost excitations are defined as positive(negative) frequency solutions with negative(positive) norm. Thus, with the assumption that $\hat \alpha > 0$, we conclude that there are ghosts in the flat sector for $\hat\beta < 0$. Although in this case there is no obvious violation of unitarity bounds (recall that $A^{(0)}$ has scaling dimension one), the mere fact that the symplectic product \eqref{ip flat 1} is not positive definite is an indication that such ghosts might occur.

Let us now focus on the case of Poincar\'e coordinates. As discussed in section \ref{sec: spec PP}, with our choice of boundary conditions at the Poincar\'e horizon, the flat sector largely resembles that of global $AdS_3$. Carrying out a calculation analogous to the one above we find that the symplectic product for flat modes in the Poincar\'e patch is given by (\ref{ip flat 3}), with the replacement of the Kronecker-$\delta$ by a Dirac $\delta$-function since $k$ is no longer quantized.

\subsection{Massive sector in global AdS}
\label{ip massive global}

Next we  evaluate the symplectic products \eqref{ip} for the positive frequency modes found in section \ref{sec: spec global adS}. We first focus on the non-flat sector, and at the end of this section we consider the hybrid boundary conditions which introduce a mixing with the flat sector. We choose to evaluate the symplectic product on a surface $\Sigma$ in which $t = {\rm const}$, in which case we obtain
\begin{equation}\label{ip global nf1}
    (A_1, A_2) = i \hat{\alpha} \int dz dx\, \varepsilon^{t \lambda \nu} \delta_1 B^*_{ \lambda} \delta_2 B_{ \nu}\, .
\end{equation}
\noindent Using the mode decomposition $\delta B_\mu = e^{\frac{i}{L}(-\omega t + k x)} b_\mu (k) $ in \eqref{ip global nf1} and computing the integral over $x$, we get
\begin{equation}\label{ip global}
    (A_1,A_2) = - 2 \pi i\hat{\alpha}\, \delta_{k_1, k_2} e^{i\frac{t}{L} (\omega_1- \omega_2)} \int_0^\infty \left( b^*_{1 \rho} b_{2 x} - b^*_{1 x} b_{2 \rho}\right) d \rho\, .
\end{equation}
\noindent It will prove convenient to express \eqref{ip global} in terms of $b_u$ and $b_v$. To do so, we recall that $b_x = \frac{1}{2}(b_u + b_v)$ along with the fact that the first order equation for $b$ yields
\begin{equation}\label{elim brho}
    b_\rho = \frac{i \alpha \rho}{ 2 k(1+\rho^2)} (b_u - b_v) - \frac{i}{2 k} (b_u + b_v)' .
\end{equation}
\noindent Therefore, we have
\begin{align}\label{brho bx}
\nonumber
    - i \int_0^\infty d \rho \left(b^*_{1 \rho} b_{2 x} - b^*_{1 x} b_{2 \rho}\right)
     ={}&
      \frac{\alpha}{2 k_1}\Bigl(\langle b_{1v}, b_{2v} \rangle - \langle b_{1u}, b_{2u} \rangle\Bigr)
      + \frac{1}{4 k_1}\Bigl[ \bigl(b_{1 u} + b_{1 v}\bigr)\bigl(b_{2 u} + b_{2 v}\bigr) \Bigr] \bigg| ^\infty_0
       \\
\nonumber
    ={}&
     \frac{\alpha \rho(1+ \rho^2)}{2 k_1(\omega_1 - \omega_2)} \Bigl[\bigl(b_{1v} b'_{2v} - b_{2v} b'_{1v}\bigr) - \bigl(b_{1u} b'_{2u} - b_{2u} b'_{1u}\bigr)  \Bigr] \bigg| ^\infty_0
      \\
    &
    + \frac{1}{4 k_1}\Bigl[ \bigl(b_{1 u} + b_{1 v}\bigr)\bigl(b_{2 u} + b_{2 v}\bigr) \Bigr] \bigg| ^\infty_0\, .
\end{align}
\noindent Here, $\langle \cdot, \cdot \rangle$ is the Sturm-Liouville (SL) product defined in appendix \ref{SL gen}. It is straightforward to verify that regularity of the modes at the origin guarantees that the contribution to \eqref{brho bx} from $\rho = 0$ vanishes, so the solutions found in \ref{sec: spec global adS} are indeed normalizable, as promised. For generic frequencies $\omega_1$ and $\omega_2\,$, the contribution from $\rho = \infty$ is finite and it evaluates to
\begin{equation}\label{IP gen omega}
    - i \int_0^\infty d \rho \bigl(b^*_{1 \rho} b_{2 x} - b^*_{1 x} b_{2 \rho}\bigr) = \frac{b^{(+)}_{2 u} b^{(-)}_{1 v} - b^{(+)}_{1 u} b^{(-)}_{2 v} }{2(\omega_1 - \omega_2)}\, .
\end{equation}
\noindent It is not hard to see that \eqref{IP gen omega} vanishes for Dirichlet, Neumann and mixed boundary conditions if $\omega_1 \neq \omega_2$. Using this fact in \eqref{ip global} we conclude that the inner product is conserved (i.e. independent of $t$) for all of the above boundary conditions, in agreement with our analysis of the symplectic flux. In order to calculate the norms, we take the limit $\omega_2 = \omega_1$ in \eqref{IP gen omega} and set $\omega_1$ to its quantized value at the end of the calculation. Since \eqref{IP gen omega} vanishes for $\omega_1 \neq \omega_2$, we can write
\begin{equation}\label{IP gen omega2}
    - i \int_0^\infty d \rho \bigl(b^*_{1 \rho} b_{2 x} - b^*_{1 x} b_{2 \rho}\bigr) = \delta_{\omega_1, \omega_2}
    \frac{1}{2} \left(b^{(+)}_{1 u} \partial_{\omega_1} b^{(-)}_{1 v} - b^{(-)}_{1 v} \partial_{\omega_1} b^{(+)}_{1 u} \right)\, .
\end{equation}
\noindent Plugging \eqref{IP gen omega2} into \eqref{ip global} we find the following general expression for the symplectic products:
\begin{equation}\label{ip global formal}
    (A_1,A_2) = \pi \hat{\alpha} \delta_{\vec{k}_1, \vec{k}_2} \left(b^{(+)}_{1 u} \partial_{\omega_1} b^{(-)}_{1 v} - b^{(-)}_{1 v} \partial_{\omega_1} b^{(+)}_{1 u} \right).
\end{equation}
\noindent We now specialize to the various boundary conditions of interest. For Dirichlet boundary conditions, the spectrum of eigenfrequencies is given by \eqref{omega D2}, \eqref{omega D zero modes}. The positive frequency solutions can be expressed more succinctly as
\begin{equation}\label{omega D positive}
   \omega^{+}_{n,k} =  2 \bigl[n + \theta(-k)\bigr] + |k| + \alpha \qquad  n = 0,1,2,\ldots\,,
\end{equation}
\noindent where $\theta(x)$ is the Heaviside function, defined as $\theta(x) = 1 $ for $x \geq 0$ and $\theta(x) = 0$ for $x < 0$. Using \eqref{omega D positive} in \eqref{ip global formal} we find
\begin{equation}\label{ip D global}
    (A_1,A_2) = \pi \hat{\alpha}  \frac{(-1)^n \pi  \alpha  \csc(\pi  \alpha) n!\, \Gamma \bigl(2\theta(-k)+|k|+n \bigr) \Gamma\bigl(- s(-k) -n-\alpha\bigr)}{4 \Gamma\bigl(1-\alpha\bigr)^2\, \Gamma\bigl(1+|k|+n+\alpha\bigr)}\, ,
\end{equation}
\noindent where $n$ is a non-negative integer. Here we have normalized the modes in such a way that the leading term has coefficient $1$. We shall continue to do so henceforth, unless explicitly otherwise stated. Similarly, the spectrum of positive frequency solutions for Neumann boundary conditions can be written as
\begin{equation}\label{omega N positive}
    \omega_N =  2 \bigl[n + \theta(k)\bigr] + |k| - \alpha \qquad  n = 0,1,2,\ldots
\end{equation}
\noindent as it follows from \eqref{omega N2} and \eqref{omega N zero modes k <0}. Inserting \eqref{omega N positive} in \eqref{ip global formal} we conclude that the Neumann norms read
\begin{equation}\label{ip N global}
    (A_1,A_2) = \pi \hat{\alpha}  \frac{(-1)^{n+1} \pi  \alpha  \csc(\pi  \alpha) n! \Gamma \bigl(2\theta(k)+|k|+n \bigr) \Gamma\bigl(-s(k) -n+\alpha\bigr)}{4 \Gamma\bigl(1+\alpha\bigr)^2\, \Gamma\bigl(1+|k|+n-\alpha\bigr)}\, ,
\end{equation}
\noindent where $n$ is a non-negative integer. Inspecting \eqref{ip D global} and \eqref{ip N global}, we note that all the modes have positive norm with the exception of the Neumann modes characterized by $n=0$, $k<0$, whose frequencies are given by \eqref{omega N zero modes k <0}. Therefore, we conclude that the theory contains ghosts for Neumann boundary conditions, as expected.

\begin{figure}[htbp]
\begin{center}
\includegraphics[scale=0.75]{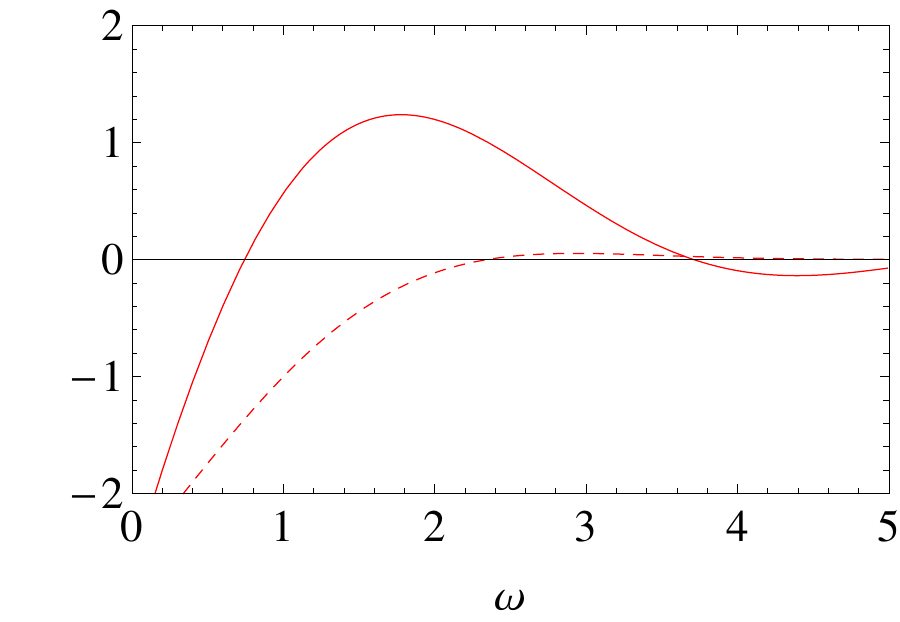}
\caption{For $\{\alpha = 0.7,\, k = -2,\, \beta = 1.1\}$, we plot the formal expression for the norm \eqref{ip global formal} with a red dashed line and the left hand side of \eqref{bc eq for w}, whose zeroes correspond to the allowed frequencies, with a red solid line. Note that the smallest frequency corresponds to a ghost. Exploring the norm numerically for various values of the parameters, we find that this behavior is generic.}
\label{ghostsmixbc}
\end{center}
\end{figure}

Let us now focus on mixed boundary conditions. In this case, the lack of a closed expression for the frequencies prevents us from displaying the norm explicitly. However, we find substantial evidence that ghosts must be present in the system for generic values of the deformation parameter $\beta$. For $\beta < 0$, the existence of ghosts follows immediately from the existence of complex frequency solutions, see for example \cite{Andrade:2011dg}. The argument is as follows. First we recall that, as all the parameters are real, the complex frequencies always occur in pairs of complex conjugate values;  c.f. figure \ref{w mix bc k neg}. Second, denoting the aforementioned solutions by $\psi_1$, $\psi_2$, we can verify that $(\psi_1, \psi_1) = (\psi_2, \psi_2) = 0$. A simple way to see this is to note that the norms have the overall time-dependent factor $\exp(- 2 t\, \mbox{Im}(w))$. Since this is in conflict with conservation, the norms must vanish. Third, the definition of the norm guarantees that cross-terms satisfy $(\psi_1, \psi_2) = (\psi_2, \psi_1)^*$, and we can explicitly verify that they are non-vanishing. Finally, diagonalizing the symplectic structure we find that one of the excitations has negative norm, signaling the presence of ghosts. We turn now to the case $\beta > 0$. In this situation we did not find evidence of complex frequency solutions, so we need to examine the norms in more detail. Indeed, we found numerical evidence that ghosts should be present for this case as well, c.f. figure \ref{ghostsmixbc}.

Finally, we consider the hybrid boundary conditions. Recall that in section \ref{sec: spec global adS} we found that there is a pair of complex frequency solutions if the absolute value of the deformation parameter $\kappa$ is large enough. Following the conventions in section \ref{sec: spec global adS}, we denote the critical value by $|\kappa_c|$. As mentioned above, these complex frequency solutions correspond to ghost/anti-ghost pairs, so we do not discuss this case any further. Let us now examine the norms of the real frequency solutions. We expect the pair of complex frequency solutions that occur for $|\kappa| > |\kappa_c|$ to remain as a ghost/anti-ghost pair when we tune $|\kappa|$ below $|\kappa_c|$. We will exhibit ample numerical evidence that this is indeed the case, and thus conclude that ghosts are present for all values of $\kappa$.

\begin{figure}[htb]
\center
\subfigure[][]{
\label{ghost hyb bc 1}
\includegraphics[width=0.42\linewidth]{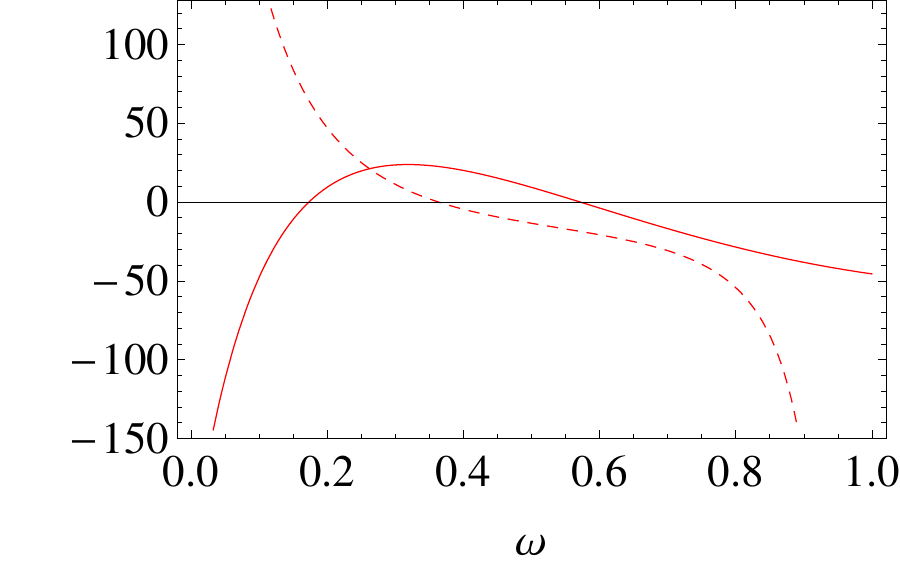}
} \qquad \qquad
\subfigure[][]{
\label{ghost hyb bc 2}
\includegraphics[width=0.42\linewidth]{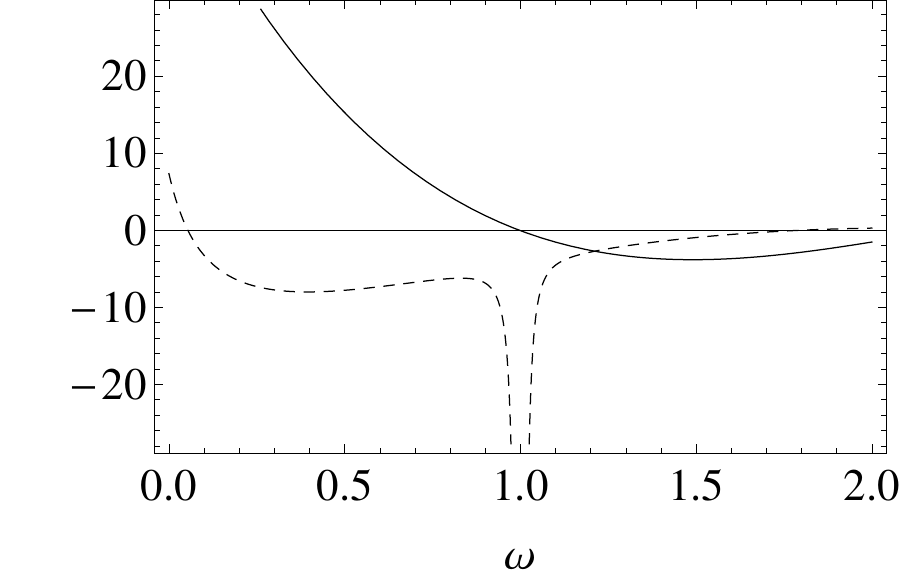}}
\caption{\ref{ghost hyb bc 1}: For $\{\alpha = 0.8,\,  k = 1,\, \kappa = 0.9\}$ we plot the left hand side of \eqref{hyb bc coeff} (solid line), whose zeros correspond to the allowed frequencies, and the expression for the norm \eqref{IP hyb final} (dashed line). We notice that the second to lowest frequency solution is a ghost. By slightly increasing the value of $\kappa$ the solutions move to the complex plane, as seen in figures \ref{tach hyb real} and \ref{tach hyb cplex}. \ref{ghost hyb bc 2}: For $\{\alpha = 0.8$,  $k = 1$, $\kappa = 0.1\}$, we plot the left hand side of \eqref{hyb bc coeff} (solid line) and the expression for the norm \eqref{IP hyb final} (dashed line). We observe that the lowest frequency mode found for higher values of $|\kappa|$ disappears, but there is still a ghost in the system. }
\end{figure}

Recall that the symplectic structure splits into the contributions from the flat and non-flat sectors as in \eqref{omega recap}, and that the symplectic product is given in terms of the symplectic structure by \eqref{ip}. Choosing the Cauchy slice $\Sigma$ on which we evaluate the product to be a surface of constant $t$, we can write the inner product as
\begin{equation}\label{ip hyb bc}
    (A_1, A_2) = (A_1, A_2)_{nf} + (A_1, A_2)_{f}\, ,
\end{equation}
\noindent where
\begin{align}\label{ip nf exp}
    (A_1, A_2)_{nf}
    ={}&
     - i \hat{\alpha} \int_{\Sigma} d^{2}x\Bigl(\delta B^*_{1 \rho} \delta B_{2x} - \delta B^*_{1x} \delta B_{2 \rho}\Bigr)
\\
\label{ip f exp}
    (A_1, A_2)_{f}
    ={}&
     i \hat{\alpha} \int_{\Sigma}d^{2}x \Bigl(\delta_1 A^{(0) *}_{ \rho} \delta_2 A^{(0)}_{ x} - \delta_1 A^{(0) *}_{ x} \delta_2 A^{(0)}_{\rho}\Bigr).
\end{align}
\noindent Next, we introduce Fourier decompositions as $\delta B_\mu = e^{\frac{i}{L}(-\omega t + k x)} b_\mu (k) $, $\delta A^{(0)}_\mu = e^{\frac{i}{L}(-\omega t + k x)} a^{(0)}_\mu$ and proceed as above by computing \eqref{ip nf exp} and \eqref{ip f exp} for generic frequencies $\omega_1$ and $\omega_2$. The expression \eqref{ip nf exp} is then given by \eqref{IP gen omega} and it only remains to compute \eqref{ip f exp}. We manipulate \eqref{ip f exp} noting that the flatness condition implies that the modes satisfy
\begin{equation}\label{flat fourier 1}
    a^{(0)}_\rho = - \frac{i}{k}\left( a^{(0)}_x\right)'\, ,
\end{equation}
\noindent where the prime denotes a radial derivative. Using \eqref{flat fourier 1} in \eqref{ip f exp} we thus find
\begin{equation}\label{ip flat fourier}
    (A_1, A_2)_f = \left. - 2 \pi \hat{\alpha}\, \delta_{k_1, k_2}\, k^{-1} a^{(0)}_{1 x} a^{(0)}_{2 x} \right|^{\rho = \infty}_{\rho = 0}\, .
\end{equation}
\noindent From the regularity condition $a^{(0)}_x =0$ at $\rho  = 0$ we conclude that only the term at $\rho = \infty$ contributes to \eqref{ip flat fourier}. Thus, gathering the results \eqref{IP gen omega} and \eqref{ip flat fourier} we find for generic frequencies
\begin{equation}\label{IP hyb generic}
    (A_1, A_2) = 2 \pi \hat{\alpha}\, e^{it(\omega_1 - \omega_2)} \delta_{k_1, k_2} \left [ \frac{b^{(+)}_{2 u} b^{(-)}_{1 v} - b^{(+)}_{1 u} b^{(-)}_{2 v}  }{2(\omega_1 - \omega_2)} - \frac{\bigl(a^{(0)}_{1 u} + a^{(0)}_{1 v}\bigr) \bigl(a^{(0)}_{2 u} +  a^{(0)}_{2 v}\bigr) }{4 k} \right].
\end{equation}
\noindent We can readily verify that the symplectic structure is conserved by noting that \eqref{IP hyb generic} vanishes for $\omega_1 \neq \omega_2$ when the boundary conditions \eqref{gen hyb}, \eqref{hyb on b}, hold. Therefore, the symplectic product can be written in terms of the coefficients of the asymptotic expansion as
\begin{equation}\label{IP hyb final}
    (A_1, A_2) = \pi \hat{\alpha}\, \delta_{\vec{k}_1, \vec{k}_2} \biggl[ b^{(+)}_u \partial_{\omega_1} b^{(-)}_{v} - \kappa^2 \left( \frac{k+\omega}{k-\omega} b^{(+)}_u \partial_{\omega_1} b^{(+)}_u  + \frac{2 k}{(k-\omega)^2} (b^{(+)}_u)^2 \right) \biggr],
\end{equation}
\noindent where $\omega$ is implicitly given by the solutions of \eqref{hyb bc coeff}. Studying \eqref{IP hyb final} numerically, we find that there is always a ghost among the lowest real frequency modes that occur for $|\kappa| < |\kappa_c|$; see figures \ref{ghost hyb bc 1}, \ref{ghost hyb bc 2}. Furthermore, we find that this picture continues to hold true for all $|\kappa|$ in the range $0 \leq |\kappa| < |\kappa_c|$. This must indeed be the case since $\kappa = 0$ corresponds to Neumann boundary conditions, which were found above to induce violations of unitarity in the bulk.

\subsection{Massive sector in Poincar\'e AdS}\label{sec ip PP}
We now proceed to compute the symplectic product for the Poincar\'e AdS modes, focusing on the non-flat piece of the connection. We start with the timelike modes. It should be noted that, since the spectrum is continuous, the products should be understood in the sense of distributions. As usual, in order to evaluate \eqref{ip} we choose $\Sigma$ to be a surface of $t = {\rm const}$, so we have
\begin{equation}\label{ip nf1}
    (A_1, A_2) = i \hat{\alpha} \int dz dx\, \varepsilon^{t \lambda \nu} \delta_1 B^*_{ \lambda} \delta_2 B_{ \nu}\,.
\end{equation}
We find it convenient to use the mode decomposition $    \delta B_\mu = e^{i(-\omega t + k x)} b_\mu (k)$ in \eqref{ip nf1} and computing the integral over $x$ we get
\begin{equation}\label{ip nf2}
    (A_1, A_2) = 2 \pi i \hat{\alpha}\, \delta(k_{1} - k_{2}) e^{i t (\omega_{1} - \omega_{2})} \int_0^\infty dz \bigl(b^*_{1 z} b_{2 x} - b^*_{1 x} b_{2 z} \bigr),
\end{equation}
\noindent where we have used $\varepsilon^{ztx} = - 1$, in consistency with the convention $\varepsilon^{zuv} = - 1$ employed in appendix \ref{sec: soln PP}.
It is convenient to express \eqref{ip nf2} in terms of the components $b_u$ and $b_v$. To this end we note that the first order equation for $b$ implies
\begin{equation}\label{bz in bu bv}
    b_z = -\frac{i \alpha}{2 k z} (b_u - b_v) - \frac{i}{2 k} \left(b_u + b_v\right)'\, ,
\end{equation}
\noindent and we recall $ b_x = \frac{1}{2}(b_u + b_v)$, $b_t = \frac{1}{2}(b_u - b_v)$. It follows that we can write
\begin{equation}\label{rad int PP}
    i \int_0^\infty dz (b^*_{1 z} b_{2 x} - b^*_{1 x} b_{2 z} ) = \frac{\alpha}{2 k} \int dz\, z^{-1}\bigl(b^*_{1 v} b_{2 v} - b^*_{1 u} b_{2 u}\bigr) - \frac{1}{4 k} \Bigl[ \bigl(b_{1 u} + b_{1 v}\bigr)^*\bigl(b_{2 u} + b_{2 v}\bigr) \Bigr] \Big|_{0}^\infty\, .
\end{equation}
\noindent The first two terms in \eqref{rad int PP} correspond to the SL inner product associated to \eqref{decoupled v} and \eqref{decoupled u}, respectively. Thus, from the results of appendix \ref{SL gen} it follows that
\begin{align}\label{rad int PP Wrons}
\nonumber
    i \int_0^\infty dz \bigl(b^*_{1 z} b_{2 x} - b^*_{1 x} b_{2 z}\bigr)
     ={}&
      \frac{\alpha z^{-1}}{2 k (m_1^2 - m_2^2)} \Bigl[ \bigl(b_{1 v} b'_{2 v}- b_{2 v} b'_{1 v}\bigr)  - \bigl( b_{1 u} b'_{2 u} - b_{2 u} b'_{1 u} \bigr) \Bigr] \Big|_{0}^\infty
       \\
     &
     - \frac{1}{4 k} \Bigl[ \bigl(b_{1 u} + b_{1 v} \bigr)^*\bigl(b_{2 u} + b_{2 v}\Bigr)] \Big|_{0}^\infty\, ,
\end{align}
\noindent where we have used the explicit form of the SL coefficients \eqref{SL bu} and \eqref{SL bv}. Next, using the near-boundary expansion one can readily verify that the contribution from $z = 0$ to \eqref{rad int PP Wrons} vanishes for Dirichlet, Neumann and mixed boundary conditions. The contribution at the Poincar\'e horizon can be evaluated by introducing a regulator $z_\infty$ at large $z$ and using
\begin{equation}\label{J large x}
    J_\nu(x) \to \sqrt{\frac{2}{\pi x}} \cos \left( x - \frac{\nu  \pi }{2} - \frac{\pi }{4} \right) \quad\mbox{for }\,\, x \gg 1 \, .
\end{equation}
\noindent From this point on the calculation proceeds in close analogy to that for a scalar field in Poincar\'e AdS. We refer the reader to \cite{Andrade:2011dg} for details.\footnote{The present calculation exhibits one additional complication: roughly speaking, the third term in \eqref{rad int PP Wrons} has the structure $\sim (m_1 m_2)^{-1/2} z_\infty \cos[(m_1 - m_2) z_\infty]$, so it is indeed power-counting divergent as $z_\infty \rightarrow \infty\,$. However, integrating this against test functions $f(m_1)$ and $f(m_2)$ of compact support, one can show that the contribution from this type of terms vanishes as we remove the regulator.} Up to terms that vanish in the distributional sense, we find that the general expression for the inner product \eqref{ip nf2} reads
\begin{equation}\label{ip PP gen}
    (A_1,A_2) = 2 \pi \hat{\alpha}\, \delta^{(2)}(k_1^i - k_2^i) {\cal Q}(\alpha, k)\, ,
\end{equation}
\noindent where
\begin{equation}\label{Q PP gen}
   {\cal Q}(\alpha, k_i) =  2 \alpha \left|A(\vec{k}) + e^{i \pi \alpha} C(\vec{k}) \right|^2\, .
\end{equation}
\noindent Here we have used that $C$ and $A$ satisfy the relation \eqref{mix bc}. Clearly, the norm \eqref{ip PP gen} is manifestly positive definite for Dirichlet, Neumann and mixed boundary conditions.

Let us now calculate the products of the spacelike excitations that are present for mixed boundary conditions. As discussed in section \ref{sec: spec PP}, their radial profile is given by \eqref{soln sl PP} with $C(\vec{k}) = 0$, which ensures normalizability since they vanish exponentially at the horizon. Furthermore, the mixed boundary condition holds provided the frequencies satisfy \eqref{mix bc tach PP}. Recall also that for all $\beta < 0$ the spectrum contains a pair of solutions $\psi_1$, $\psi_2$ whose frequencies are complex conjugate to each other. As argued above, there is always a ghost among these degrees of freedom, so we do not consider this case any further. For $\beta >0$ we found real frequency spacelike solutions, whose norm we now compute.

Since the real frequency spacelike solutions form a discrete set, we compute their norms in analogy to the calculation of the inner product in global coordinates. Our starting point is the general expression \eqref{rad int PP Wrons}. Because the radial profiles decay exponentially at the horizon, the only non-vanishing contribution to \eqref{rad int PP Wrons} comes from the boundary asymptotics. A simple computation reveals that the norm of the real frequency spacelike solutions can be written in terms of the coefficients of the asymptotic expansion as
\begin{equation}\label{norm tach1}
    (A_1, A_2)_{T} = \pi \hat{\alpha}\, \delta(k_1 - k_2) \Bigl(b^{(+)}_u \partial_\omega b^{(-)}_v - b^{(-)}_v \partial_\omega b^{(+)}_u\Bigr)\, ,
\end{equation}
\noindent where $\omega$ satisfies \eqref{mix bc tach PP}. Plugging in \eqref{norm tach1} the explicit expressions for the coefficients $b^{(\pm)}$ found previously in \eqref{bplus tach} and \eqref{bminus tach}, we find that the norm of the spacelike solution is
\begin{equation}\label{norm tach2}
    (A_1, A_2)_{T} =  \pi \hat{\alpha}\, \delta(k_1 - k_2)  |A|^2 \frac{\pi \alpha (k- \alpha \omega) \csc (\pi  \alpha)}{p^2}\, .
\end{equation}
\noindent Note that the sign of the norm \eqref{norm tach2} is controlled by the factor $(k - \alpha \omega)$. Now, it follows from \eqref{mix bc tach PP} that positive norm solutions occur for positive and negative frequencies, see also figure \ref{w real tach}. Thus, we conclude that there are ghosts in the theory for mixed boundary conditions and $\beta > 0$.

It only remains to discuss hybrid boundary conditions. In this case, the spectrum consists of both real and imaginary frequencies, regardless of the value of the deformation parameter $\kappa$. As argued above, the existence of non-real frequencies is associated with ghosts on general grounds. Therefore, no detailed calculation of the norms is required to show that this class of theories violate unitarity in the bulk.

As pointed out in \cite{Andrade:2011dg}, bulk theories dual to CFT's in which the unitarity bound is violated do not necessarily contain ghosts when the geometry is that of Poincar\'e AdS. Alternatively, the two-point function suffers from a divergence near the light-cone, which implies that the theory does not exist. This motivates us to inspect the near light-cone structure of the Neumann correlators.

The boundary (Wightman) two-point function can be easily computed given the matrix of symplectic products, see e.g. \cite{Andrade:2011dg}. For the timelike modes in the Neumann theory we find
\begin{equation}\label{2pt N}
    \langle 0| b^{(+)}_u (-k_i) b^{(+)}_u (k_i) |0 \rangle = (A_1,A_2)^{-1} |_{Neumann} = \frac{4^\alpha q^2 L}{\pi \alpha^2 \Gamma(-\alpha)^2}  \frac{(\omega-k)^{1-\alpha}}{(\omega+k)^{1+\alpha}}\, .
\end{equation}
\noindent In order to obtain \eqref{2pt N} we have normalized the radial profiles such that the leading term is $1$. As expected, the Fourier transform does not converge due to the behavior near $\omega = -k$; this behavior is to be contrasted with the Dirichlet case, in which we find
\begin{equation}\label{2pt D}
    \langle 0| b^{(-)}_v (-k_i) b^{(-)}_v (k_i) |0 \rangle = (A_1,A_2)^{-1} |_{Dirichlet} =   \frac{4^{-\alpha} q^2 L}{\pi \alpha^2 \Gamma(\alpha)^2}\frac{(\omega+k)^{1+\alpha}}{(\omega-k)^{1-\alpha}}\, .
\end{equation}
\noindent This is clearly finite as we approach $\omega = - k$. In the parameter range of interest, namely $0 < \alpha < 1$, the divergence near $\omega = k$ is mild enough so that the Fourier transform of \eqref{2pt D} converges.

The parallel with \cite{Andrade:2011dg} extends beyond the existence of the light-cone divergence discussed above, in that this divergence can be related to the appearance of lightlike gauge modes. In fact, recall that in section \ref{sec: spec PP} we found that the Neumann theory admits the lightlike solution \eqref{RM Neumann}, which in position space can be written as
\begin{equation}\label{RM Neumann gen}
    \delta B = f(u) z^{-\alpha}  du\, ,
\end{equation}
\noindent where $f$ is an arbitrary function of $u$. From \eqref{ip nf2}, it is clear that the aforementioned solution has zero norm since its $z$-component vanishes. Moreover, it is straightforward to verify that the inner product of \eqref{RM Neumann gen} with the timelike modes vanishes in the distributional sense. Thus, assuming that the spectrum of the Neumann theory we found in \ref{sec: spec PP} is complete,\footnote{In principle, there could be solutions with anharmonic time dependence, which lie outside of the class of modes we consider here. Although we have not studied this possibility in detail, the present setup is self-consistent in that it provides the correct physical results, namely that the Neumann theory is sick.} i.e. that there are only timelike and lightlike modes, we conclude that the lightlike solution \eqref{RM Neumann gen} is a null direction of the symplectic structure and is thus pure gauge. The reader might be somewhat puzzled by the fact that there is a gauge mode which is not flat. However, one can argue that this must be the case by noting the large arbitrariness in \eqref{RM Neumann gen} parametrized by the function $f(u)$, which is unconstrained by the equations of motion.

\section{Discussion}\label{section:discussion}
By studying the bulk symplectic structure, we have obtained a class of admissible boundary conditions for the MCS system in asymptotically-AdS spaces. According to the holographic dictionary, these boundary conditions determine the operator content in the possible dual theories. In agreement with the existing literature, we find that there is a vector operator of conformal dimension $1 \pm \alpha\,$, in addition to the well-known chiral currents which are also present in the pure Chern-Simons theory. The vector operator is associated with the Hodge dual of the bulk field strength, which behaves as a massive vector with a mass proportional to the Chern-Simons coupling. It is worth mentioning that the components of these operators satisfy a constraint, so they have less degrees of freedom than naively expected. This feature is reminiscent of the situation in topologically massive gravity, where similar constraints exist \cite{Skenderis:2009nt}. The chiral currents, on the other hand, are associated to the flat piece of the connection, and are in that sense topological.

Our analysis reveals that, whereas it is possible to impose boundary conditions such that the topological and massive sectors decouple, it is also in principle valid to introduce a mixing between them. In particular, we studied a class of boundary conditions that corresponds to double-trace deformations that couple the chiral currents with the vector operators. Regarding the boundary conditions in which both sectors decouple, we have also considered boundary conditions that yield double-trace deformations within each sector. In this case, it is even possible to generalize these to incorporate multi-trace deformations in the usual way. Our main result is that this apparently large freedom in the choice of boundary conditions is severely restricted once we impose unitarity as an extra requirement.

We have addressed the issue of unitarity by studying the MCS theory both in Poincar\'e and global AdS. In these setups, the violations of unitarity generically manifest themselves as ghost excitations in the spectrum of the theories defined with given boundary conditions. The boundary conditions that pass the test of unitarity correspond to fixing the leading behavior of the massive sector (Dirichlet boundary conditions), while separately specifying a linear relation between the two components of the flat connection along the boundary directions. It is worth mentioning that the latter also requires a specific choice of sign in the proportionality constant. Furthermore, we mention that for boundary conditions that fix the spatial part of the boundary connection, the topological degrees of freedom become pure gauge (in the absence of holonomies).

For the boundary conditions corresponding to double-trace deformations which involve the massive sector, we contented ourselves with numerical results and the reader may wonder whether our analysis was exhaustive enough to rule out the existence of a non-trivial phase diagram. In particular, since the Dirichlet theory is well defined and we can in principle approach it by continuously tuning the deformation parameters, it is reasonable to ask whether there is an open set of unitarity-preserving values near the Dirichlet point. The answer to this question is negative, as it is most easily seen when the geometry is taken to be the Poincar\'e patch of $AdS_{3}$. In this case, the presence of a scaling symmetry dictates that, up to sign changes, all non-zero values of the coupling constants are equivalent. One can use this fact to draw conclusions regarding the mixed boundary condition $\delta B^{(+)}_u = \beta\, \delta B^{(-)}_v$. For the reason we just mentioned, it suffices to study the cases $\beta = 0,\, \infty,\, \pm 1$, where $\beta = 0 $ corresponds to Dirichlet boundary conditions. Then, finding ghosts for $\beta = \pm 1$ implies that these remain for all non-zero $\beta$, forbidding a non-zero critical value. Moreover, noting that the Poincar\'e patch theory captures the high-momentum dynamics of the theory in global AdS, \footnote{This is most easily seen by noting that, for short characteristic lengths, a cylinder is equivalent to a plane.} one can extend this result to the global case. Clearly, the analogous statement holds true for our hybrid boundary conditions parametrized by the constant $\kappa$.

In many scenarios, the presence of the ghosts is in one-to-one correspondence with violations of the unitarity bound in the dual theory, which establishes that the scaling dimension of vector operators must be greater than one. In fact, the operator of dimension $1-\alpha$ violates the bound for all $\alpha$ and, accordingly, we find ghosts whenever the corresponding slow-decaying branch fluctuates. The only exception are the conformally invariant Neumann boundary conditions in the Poincar\'e patch, which set to zero the faster fall-off. In analogy with the scalar case discussed in \cite{Andrade:2011dg}, we have found in this case that the spectrum is ghost-free, and that the expected pathologies arise instead in the 2-point function, which is ill-defined even at large distances. Interestingly, we also found ghosts in the flat sector, which occur for some choices of the parameter that controls the double-trace deformation. Since the chiral currents have dimension one, these unitarity violations cannot be linked to the bound on the scaling dimension. These pathologies are indeed to be expected, however, because the expression for the symplectic product restricted to the flat sector is not positive definite in any obvious way.

It is worth commenting in more detail on the mixed boundary conditions in relation to the unitarity bound. Above, we obtained these as a deformation of the Neumann theory, as it is customary for bulk scalars whose mass lies in the Breitenlohner-Freedman window. Had the Neumann theory been well defined, the inclusion of the relevant double-trace operator could have been thought of as triggering an RG flow towards the Dirichlet theory. However, as we have seen, the Neumann theory is sick, and inclusion of the double-trace operator does not cure its pathologies. Thus, the aforementioned flow is not well defined. Attempting to remedy this, one might try to understand the mixed boundary conditions as triggering a flow from the Dirichlet theory. In this case, unfortunately, the deformation term one needs to add is of the form $\sim \Bigl(B^{(-)}_v\Bigr)^2$, so it corresponds to an irrelevant operator of dimension $2(1+\alpha)$. It follows that the resulting theory is non-renormalizable and the ghosts that arise can be understood as being the result of our loss of control of the theory in the UV.

It is interesting to contrast our results with those of \cite{DHoker:2010hr}, in which the authors found an effective three-dimensional MCS theory in the context of holographic RG flows. More precisely, they constructed five-dimensional solutions in the Einstein-Maxwell-Chern-Simons theory which have the interpretation of magnetic branes. Perturbations around these backgrounds turn out to describe RG flows from four-dimensional field theories in the UV to two-dimensional ones in the IR, and the dynamics of the latter are captured by a 3$d$ MCS theory. Requiring the usual Dirichlet boundary conditions in the UV and imposing matching conditions in the bulk, the effective IR theory contains a double-trace for the vector operators which we denoted by $B^{(\pm)}$. Our analysis reveals that this theory must contain ghosts, and indeed, the results of \cite{DHoker:2010hr} indicate that violations of unitarity are present if one tries to extend the domain of validity of the IR description to the entire bulk. Then, what saves the theory is the existence of an effective cut-off associated to the domain wall solution, whose presence implies that the IR description breaks down at some intermediate value of the radial coordinate. This is to be expected since Dirichlet boundary conditions were imposed in the UV, and these respect the dual unitarity bounds. The issue of removing the bulk ghosts by introducing the appropriate cut-offs will be discussed in an upcoming publication
\cite{T.Andrade}.

We now briefly comment on the implications of our results in the context of potential condensed matter applications. For illustrative purposes, we first review the relevant results of the pure Maxwell theory and then move on to describe how the addition of the Chern-Simons term changes the picture. In terms of the radial variable of \eqref{Global AdS3}, the asymptotics of the gauge field in the pure Maxwell theory are of the form
\begin{equation}\label{maxwell asympt}
    A_i = \log r A^{(1)}_i + A^{(0)}_i + \ldots \qquad {\rm with} \quad \nabla^{(0)}_i A^{(1) i} = 0\, ,
\end{equation}
\noindent where $i$ is a boundary index and $\nabla^{(0)}$ is the covariant derivative associated with the conformal boundary metric. The conservation equation satisfied by the coefficient $A^{(1)}_i$ indicates that it should be interpreted as the $U(1)$ current. This fact was overlooked in
\cite{Ren:2010ha,Nurmagambetov:2011yt,Liu:2011fy,Lashkari:2010ak}, in which the authors discussed the construction of a holographic $1+1$ dimensional superfluid/superconductor incorrectly interpreting $A^{(0)}$ as the boundary current. We mention that this confusion was resolved in \cite{Jensen:2010em} using the conservation argument given above. However, there is still an obstruction to the study of such holographic theory, since the boundary conditions that allow for a fluctuating current yield ghosts \cite{Andrade:2011dg}. Thus, the applicability of the by now standard procedure \cite{Hartnoll:2008vx,Hartnoll:2008kx} to the study of holographic $1+1$ superconductors remains, at least, unclear. Given this, it is compelling to ask ourselves what are the implications of adding the Chern-Simons term to the Maxwell theory and the possible AdS/CMT applications of the resulting setup.\footnote{This possibility was suggested in \cite{Lashkari:2010ak}, with a different motivation.} As we have seen, the inclusion of the Chern-Simons term drastically modifies the scenario, as the $U(1)$ vector current is replaced by the topological chiral currents associated to the flat connections, and one can imagine introducing an order parameter (dual to a minimally coupled bulk charged scalar, say) which could potentially break the associated symmetry spontaneously.\footnote{We thank Per Kraus for pointing out this possibility.} We leave the exploration of this line of research for future work.

\vskip 1cm
\centerline{\bf Acknowledgments}

We are grateful to Geoffrey Comp\`ere, Eduardo Fradkin, Gary Horowitz, Per Kraus, Mukund Rangamani, Simon Ross and Jorge Santos for helpful conversations, and specially to Don Marolf for many useful discussions on these and related topics. T.A. was partly supported by a Fulbright-CONICYT fellowship, by the US National Science Foundation under grant PHY08-55415 and by funds from the University of California. T.A. is also pleased to thank the Department of Physics of the University of Illinois at Urbana-Champaign and the Department of Mathematics of the University of California, Davis, for their hospitality during the completion of this work. RGL is partially supported by the US Department of Energy under contract FG02-91-ER40709. The work of J.I.J. is supported by the research programme of the Foundation for Fundamental Research on Matter (FOM), which is part of the Netherlands Organisation for Scientific Research (NWO).

\appendix
\section{Solutions to the equations of motion on $AdS_{3}$}\label{section:solutions}
Here we consider the MCS equation \eqref{MaxCS equation} on a fixed background geometry. We remark that since the MCS system is linear, in the probe approximation both the background gauge field and its fluctuations satisfy the same equation. Splitting the gauge field fluctuation as in \eqref{definition B}, i.e. $\delta A  = \delta B + \delta A^{(0)}$, we have that $\delta A^{(0)}$ is flat and $\delta B$ satisfies
\begin{equation}\label{eom1 B}
0 = \epsilon^{\mu\nu\rho}\partial_{\nu}\delta B_{\rho} + \frac{\alpha}{L}\delta B^{\mu}\, ,
\end{equation}
\noindent where $\epsilon^{\mu\nu\rho} = -1/\sqrt{|g|}\varepsilon^{\mu\nu\rho}$ and $g$ is the determinant of the background metric. In what follows we present the general solution of this equation for the backgrounds of interest, namely $AdS_{3}$ in both the Poincar\' e patch and global coordinates.

\subsection{Poincar\'e patch of $AdS_{3}$}\label{sec: soln PP}
We first consider the background geometry to be the Poincar\'e patch of $AdS_{3}\,$, with line element given by
\begin{equation}\label{ds2}
    ds^2 = \frac{L^{2}}{z^2} \left(dz^2 - dt^2 + dx^2\right).
\end{equation}
\noindent We remind the reader that the line element \eqref{ds2} posses the symmetry
\begin{equation}\label{dilation PP}
    z \rightarrow \lambda z\, , \qquad t \rightarrow \lambda t\, , \qquad x \rightarrow \lambda x\, ,
\end{equation}
\noindent which corresponds to dilations in the dual theory. This symmetry play an important role in the analysis of the spectrum, as discussed in the main text. It will prove convenient to introduce null coordinates
\begin{equation}\label{nc}
    t = u - v\, , \quad x = u + v\, ,
\end{equation}
\noindent so that the line element reads
\begin{equation}\label{ds2 nc}
    ds^2 = \frac{L^{2}}{z^2} \left(dz^2 + 4 du dv\right).
\end{equation}
\noindent By convention, we take $\varepsilon^{z u v} = -1$. We Fourier-decompose the fluctuations of the gauge-invariant (massive) mode as
\begin{equation}\label{modes}
    \delta B_\mu(u,v;z) = e^{i(k_u u + k_v v)} b_\mu(z),
\end{equation}
\noindent where $k_u = k - \omega$ and $k_v = k+\omega$. Inserting \eqref{ds2 nc} and \eqref{modes} in \eqref{eom1 B} we find
\begin{align}\label{1st order z}
    0
     = {}&
    2 \alpha\, b_z + i z \bigl(k_u b_v - k_v b_u\bigr)\, ,
    \\
\label{1st order u}
   0 ={}&
    \alpha\, b_v + z \bigl(i k_v b_z - b_v' \bigr) \, ,
    \\
\label{1st order v}
    0 = {}&
    \alpha\, b_u + z \bigl(- i k_u b_z + b_u' \bigr)\, .
\end{align}
\noindent The above equations can be decoupled by going to second order in derivatives, obtaining
\begin{align}\label{decoupled z}
  0
   = {}&
  z^2 b''_z - z b'_z - \bigl[\alpha^2 -1 - m^2 z^2\bigr] b_z \, ,
\\
\label{decoupled u}
  0
  ={}&
  z^2 b''_u - z b'_u - \bigl[\alpha(\alpha + 2) - m^2 z^2\bigr] b_u \, ,
\\
\label{decoupled v}
  0
  ={}&
  z^2 b''_v - z b'_v - \bigl[\alpha(\alpha - 2) - m^2 z^2\bigr] b_v \, ,
\end{align}
\noindent where $m^2 \equiv - k_u k_v = \omega^2 - k^2$ is the eigenvalue of the Laplacian associated to the conformal boundary metric. Once the general solution to the second order equations has been found, the first order equations provide relations among the various integration constants.
\subsubsection{Timelike modes ($m^{2}>0$)}
For timelike momenta and $\alpha \notin \mathds{Z}$ we can write the general solution of \eqref{1st order z}-\eqref{decoupled v} as
\begin{align}
\nonumber
b_{z} ={}&
im z\left[A(\vec{k})  J_{\alpha}(mz) +C(\vec{k})  J_{-\alpha}(mz)\right]
 \\
\label{soln tl PP}
 b_{u} ={}&
  k_{u} z\left[ - A(\vec{k}) J_{1+\alpha}(mz) + C(\vec{k}) J_{-1-\alpha}(mz)\right]
  \\
\nonumber
 b_{v} ={}&
k_{v} z \left[ A(\vec{k}) J_{-1+\alpha}(mz) - C(\vec{k}) J_{1-\alpha}(mz)\right],
\end{align}
\noindent where $J_\nu$ is the Bessel function of the first kind.

\subsubsection{Spacelike modes $(m^{2}<0)$}
In this case we define $m^2\equiv - p^2$, so that $p^{2}>0$. The general solution is then
\begin{align}
\nonumber
b_{z} ={}&
i p z\left[A(\vec{k})  K_{\alpha}(p z) +C(\vec{k})  I_{\alpha}(pz)\right]
 \\
\label{soln sl PP}
 b_{u} ={}&
  k_{u} z\left[ A(\vec{k}) K_{1+\alpha}(pz) - C(\vec{k}) I_{1+\alpha}(pz)\right]
  \\
\nonumber
 b_{v} ={}&
k_{v} z \left[ A(\vec{k}) K_{-1+\alpha}(pz) - C(\vec{k}) I_{-1+\alpha}(pz)\right],
\end{align}
\noindent where $I_\nu$ and $K_\nu$ are the modified Bessel functions of the first and second kind, respectively.

\subsubsection{Lightlike modes ($m^{2}=0$)}
\begin{itemize}
\item Right-moving ($k_{v}=0$):
\begin{equation}\label{soln PP kv=0}
b_{z} = - C\,\frac{ik_{u}}{2\alpha}z^{1+\alpha}\, ,\qquad b_{u} =  Az^{-\alpha}+C\,\frac{k_{u}^2}{4\alpha(1+\alpha)}z^{2+\alpha} \,,\qquad b_{v} = C\,z^{\alpha}\, .
\end{equation}
\item Left-moving ($k_{u}=0$):
\begin{equation}\label{soln PP ku=0}
b_{z} = A\,\frac{ik_{v}}{2\alpha}z^{1-\alpha}\, ,\qquad b_{u} = A\, z^{-\alpha}\,,\qquad b_{v} = A\, \frac{k_{v}^2}{4\alpha(\alpha-1)}z^{2-\alpha} + Cz^{\alpha}\, .
\end{equation}
\end{itemize}

\subsection{Global $AdS_{3}$}\label{sec:soln global}

We now consider the global $AdS_3$ metric written as
\begin{equation}\label{Global AdS3}
ds^{2} = \frac{dr^{2}}{1 +\displaystyle{ \frac{r^{2}}{L^{2}}}} - \left(1 + \frac{r^{2}}{L^2}\right)dt^{2} + \frac{r^{2}}{L^{2}}dx^{2}\, ,
\end{equation}
\noindent where we have defined $x \equiv L\varphi$.
We will use the dimensionless radial coordinate $\rho = r/L$, so we write
\begin{equation}\label{ansatz global AdS}
\delta B = e^{\frac{i}{L}\left(-\omega t  + k x\right)}\Bigl[L\, b_{\rho}(\rho)\,d\rho + b_{t}(\rho)\, dt + b_{x}(\rho)\,dx\Bigr].
\end{equation}
\noindent Single-valuedness of the solution demands that we identify $x \sim x + 2\pi L$, so the dimensionless momentum $k$ is an integer, $k \in \mathds{Z}$. Following the same steps as before to decouple the equations we find
\begin{align}\label{eq plus}
0={}&
 \rho^2\left(1+\rho^2\right) b_u''+\rho\left(1+3\rho^2\right)b_u'+\left(\frac{\rho^2}{1+\rho^2}\omega^2 -\alpha(\alpha+2)\rho^2- k^2\right)b_{u}
\\
0={}&
 \rho^2\left(1+\rho^2\right) b_v''+\rho\left(1+3\rho^2\right)b_v'+\left(\frac{\rho^2}{1+\rho^2}\omega^2 -\alpha(\alpha-2)\rho^2- k^2\right)b_{v}\, ,
 \label{eq minus}
\end{align}
\noindent where  $b_{v}=b_{x}-b_{t}$ and $b_{u} = b_{t}+b_{x}$ as before. Notice that the equations are related by $\alpha \leftrightarrow -\alpha$. When $k \neq 0$, their solution is
\begin{align}\label{solution b in global AdS3}
b_{u} ={}&
C_{u}^{(+)}F( \omega, k,\alpha;  \rho) + C_{u}^{(-)} F(  \omega, -k, \alpha; \rho)
\\
b_{v} ={}&
C_{v}^{(+)}F(  \omega, k, -\alpha;  \rho) + C_{v}^{(-)} F( \omega, -k,-\alpha;  \rho)
\end{align}
\noindent with
\begin{equation}\label{F hyp}
    F( \omega, k, \alpha; \rho) = \rho^{k} (1 + \rho^{2})^{\omega/2}\, _{2}F_{1} \left( \frac{1}{2}(k-\alpha+\omega), \frac{1}{2}(2+k+\alpha+\omega); 1+k; - \rho^{2} \right),
\end{equation}
\noindent where $_{2}F_{1}$ is the Gauss Hypergeometric function. We can obtain $b_{\rho}$ from the $\rho$ component of \eqref{first order equation for B}, which reads
\begin{align}
0
={}&
\alpha\, b_{\rho} + \frac{i}{\rho(1+\rho^2)}\left(\omega b_{x} + k b_{t}\right)
\nonumber\\
={}&
\alpha\, b_{\rho} + \frac{i}{2\rho(1+\rho^2)}\Bigl[\left(k+\omega\right) b_{u}+\left(\omega-k \right) b_{v} \Bigr].\label{radial constraint global ads}
\end{align}
\noindent  Finally, inserting our solution back into the first order equation \eqref{first order equation for B} we find the relations
\begin{align}\label{the cs}
C_{v}^{(+)} ={}&
\frac{k-\alpha+\omega}{k+\alpha-\omega}  C_{u}^{(+)}
\\
C_{v}^{(-)} ={}&
 \frac{k+\alpha+\omega}{k-\alpha-\omega} C_{u}^{(-)}\, ,
\label{the cs 2}
\end{align}
\noindent so there are only two independent degrees of freedom.

When $k=0$, the basis of solutions \eqref{solution b in global AdS3} is no longer valid. Instead, one can use
\begin{align}\label{solution b in global AdS3 k = 0}
b_{u}
 ={}&
  (1 + \rho^{2})^{\omega/2} \left[
\tilde{C}_{u}^{(+)}\tilde{F}_1( \omega, \alpha;  \rho) + \tilde{C}_{u}^{(-)} \tilde{F}_2 (  \omega, \alpha; \rho) \right]
\\
b_{v}
 ={}&
(1 + \rho^{2})^{\omega/2}\left[\tilde{C}_{u}^{(+)} \tilde{F}_1( \omega,  -\alpha;  \rho) + \tilde{C}_{u}^{(-)}  \tilde{F}_2 ( \omega,-\alpha;  \rho) \right],
\end{align}
\noindent where
\begin{align}\label{F1 tilde}
    \tilde{F}_1( \omega, \alpha;  \rho)
     ={}&
      \, _{2}F_{1} \left( a, b; 1; - \rho^{2} \right),
\\
\label{F2 tilde}
    \tilde{F}_2( \omega, \alpha;  \rho)
     ={}&
     _{2}F_{1} \left( a , b ; 1; - \rho^{2} \right) \log(- \rho^2)
     \nonumber
      \\
            + \sum_{n=1}^{\infty}
        & \frac{(a)_n
         (b)_{n}}{(n!)^2} (-\rho^2)^n \Bigl[\psi(a+n)- \psi(a) + \psi(b+n)- \psi(b) - 2 \psi(n+1) + 2 \psi(1) \Bigr].
\end{align}
Here $a= (\omega-\alpha)/2$ , $b = (2+\alpha+\omega)/2$ and $(a)_n$ is the Pochhammer symbol defined by $(a)_n = \Gamma(a+n)/\Gamma(a)$. The presence of the logarithm in \eqref{F2 tilde} makes the solution non-normalizable at the origin. Therefore, we set $\tilde{C}_{u}^{(-)}  = \tilde{C}_{u}^{(-)} =0\,$, and will not consider the profile \eqref{F2 tilde} in the body of the paper. Once again, $b_\rho$ can be obtained using \eqref{radial constraint global ads}. The full solution is then obtained noting that the first order equations require
\begin{equation}\label{the cs tilde}
\tilde{C}_{v}^{(+)} = - \tilde{C}_{u}^{(+)}\, .
\end{equation}

\section{Sturm-Liouville problem}\label{SL gen}
The Sturm-Liouville eigenvalue problem in the interval $x \in (a,b)$ is characterized by the second order ODE
\begin{equation}\label{SL PP}
    L \psi = \lambda \psi\,, \qquad {\rm where} \quad      L = \frac{1}{w(x)}\left[ -\frac{d}{dx} \left(p(x) \frac{d}{dx} \right)  + q(x)  \right].
\end{equation}
\noindent For given boundary conditions at $x = a$ and $x = b$, the solution of the problem corresponds to a given set of eigenfunctions $\psi_\lambda$ with eigenvalue $\lambda$. We can define an associated inner product as
\begin{equation}\label{SL ip PP}
    \langle \psi_1, \psi_2 \rangle = \int_a^b dx\, w(x) \psi_1^* \psi_2\, .
\end{equation}
\noindent If the boundary conditions are such that the operator $L$ is self-adjoint, it follows that the eigenvalues are real. In addition, eigenfunctions with different eigenvalue are orthogonal with respect to \eqref{SL ip PP}. Moreover, integrating by parts in the expression $\langle L \psi_1, \psi_2 \rangle$ one can show
\begin{equation}\label{SL as wronskian}
    \langle \psi_1, \psi_2 \rangle = \frac{p(x)}{\lambda_1 - \lambda_2} \left[ \psi_1^* \frac{d}{dx} \psi_2 - \psi_2 \frac{d}{dx} \psi_1^*   \right] \bigg |_{a}^b\,\, ,
\end{equation}
\noindent which reduces the calculation of the integral in the l.h.s. to a simple expression that only involves the asymptotics of $\psi$. In order to apply \eqref{SL as wronskian} to the computation of symplectic products, we write the integrals of interest in terms of variables which satisfy decoupled equations of the form \eqref{SL PP} for some $\lambda$, $p$, $w$, $q$. In particular, we note that \eqref{decoupled u} adopts the SL form with $\lambda = m^2$ and
\begin{equation}\label{SL bu}
    p = z^{-1}\,, \qquad w = z^{-1}\,, \qquad q = \alpha(\alpha + 2) z^{-3}\, ,
\end{equation}
\noindent while for \eqref{decoupled v} we have $\lambda = m^2$ in addition to
\begin{equation}\label{SL bv}
    p = z^{-1}\, , \qquad w = z^{-1}\, , \qquad q = \alpha(\alpha - 2) z^{-3}\, .
\end{equation}
\noindent Similarly, equation \eqref{eq plus} can be written as a SL problem with $\lambda = \omega^2$ and
\begin{equation}\label{SL coeff u}
    p = \rho (1+\rho^2) \,,\qquad w =  \frac{\rho}{1+\rho^2}\,, \qquad  q = \frac{k^2}{\rho} + \alpha(\alpha + 2) \rho\, ,
\end{equation}
\noindent while for \eqref{eq minus} we have $\lambda = \omega^2$ and
\begin{equation}\label{SL coeff v}
    p = \rho (1+\rho^2)\, ,\qquad w =  \frac{\rho}{1+\rho^2}\, , \qquad  q = \frac{k^2}{\rho} + \alpha(\alpha - 2) \rho\, .
\end{equation}

\section{$U(1)$ symmetries}\label{section:sym bndy}
The goal of this appendix is to study the character of the $U(1)$ transformations from the point of view of possible boundary field theory duals. We recall that these transformations act on the fields as
\begin{equation}\label{u1}
    \delta_\lambda A_\mu = \partial_\mu \lambda \, ,
\end{equation}
\noindent where $\lambda$ is a single-valued arbitrary function of spacetime. By considering only single-valued functions we are ruling out ``large" gauge transformations, which are not connected with the identity. Our approach will be to note that symmetries of the bulk theory that have a non-trivial action on the boundary data (i.e. the boundary sources and operators) are naturally interpreted as symmetries of the boundary theory. Our analysis follows that of \cite{Barnich:2001jy,Barnich:2007bf,Compere:2008us} regarding boundary diffeomorphisms. For the sake of concreteness, we shall assume in this section that the background geometry is global AdS. Analogous results apply to the Poincar\'e case.

Quite generally, a symmetry is a transformation that preserves the phase space of the theory. As a consequence, it follows that such a transformation must leave both the action and the boundary conditions invariant. From the boundary point of view, this ensures that the quantum generating functional is invariant under the symmetry of interest. Obviously, the $U(1)$ is a symmetry of the equations of motion \eqref{MaxCS equation}. However, if this transformation acts non-trivially on the boundary, it may or may not preserve the boundary conditions and the action functional. In fact, depending on the boundary conditions, the $U(1)$ transformation could be a global symmetry, a gauge symmetry or might not even be a symmetry at all. By a gauge symmetry we mean a transformation whose generator (i.e. the associated charge) vanishes on shell. Global symmetries, on the other hand, have non-zero charge. It should be noted that this charge ought to be integrable and finite for the corresponding transformation to be properly implemented in phase space.

Let us first establish the fact that the $U(1)$ symmetries that act trivially on the boundary are pure gauge (in the sense that they are null directions of $\Omega$).  First, we note that these are in fact symmetries since they leave both the action and the boundary conditions invariant. Furthermore, by a calculation identical to the one in section \ref{ip flat sector}, we conclude that the symplectic product of $\delta_\lambda A = \nabla \lambda$ with an arbitrary configuration $\delta A$ is given by
\begin{equation}\label{gauge}
    \Omega(\delta_\lambda A, \delta A) = \hat{\alpha} \int_{\partial \Sigma} dx\, \lambda\, \delta A^{(0)}_x\, ,
\end{equation}
\noindent where $\delta A^{(0)}_x$ is the flat piece of $\delta A$ and $\partial \Sigma$ corresponds to the intersection of the boundary with a $t = {\rm const.}$ slice. Comparing \eqref{gauge} with \eqref{d Q}, we conclude that the $U(1)$ transformations that leave the boundary invariant have vanishing charge and are thus pure gauge. As mentioned above, since these transformations leave the boundary data invariant, they do not manifest in the dual theory.

We now consider transformations whose action on the boundary data is non-trivial, and examine their action on the various boundary conditions under consideration. We begin with hybrid boundary conditions. Since these mix the flat connections with the massive sector, and the $U(1)$ acts only on the former, it is clear that the $U(1)$ transformations are not symmetries of the theory. Let us now focus on boundary conditions which do not mix the flat and massive sectors. Since the $U(1)$ transformations only affect the flat piece of the connection, we concentrate our attention in the flat sector. It follows from \eqref{gauge} that theories in which $A_x^{(0)}$ is fixed have the residual gauge symmetry associated with $\lambda |_{\partial M} = \lambda(t)$. Since this acts non-trivially at the boundary, we conclude that the gauge symmetry is also present in the dual theory. Consider now the boundary condition $\delta A^{(0)}_t = 0$, in which case the residual symmetry corresponds to $\lambda |_{\partial M} = \lambda(x)$. We see from \eqref{gauge} that there is an associated non-vanishing infinitesimal charge, given by the right hand side of \eqref{gauge} (see section \ref{section:symplectic}, in particular eq. \eqref{d Q}). We notice that \eqref{gauge} is trivially integrable, so we can write an expression for the total charge as
\begin{equation}\label{Q}
    Q_\lambda = \hat{\alpha} \int_{\partial \Sigma} dx\, \lambda(x) A^{(0)}_x\, .
\end{equation}
\noindent The existence of the well-defined non-vanishing charge \eqref{Q} implies that the associated symmetry is global. Once again, this transformation acts non-trivially on the boundary so it is present in the boundary theory. It is worth emphasizing that the aforementioned symmetry is in fact infinite dimensional, and that the charges \eqref{Q} correspond to the chiral currents discussed previously in the literature. This can be seen by computing the Poisson bracket $\{Q_\lambda, Q_\sigma \}$. The easiest way to proceed is to note that the charges are the generators of the associated symmetry, so it must be the case that
\begin{equation}\label{PB 1}
    \{ Q_\lambda, Q_\sigma \} = \delta_\sigma Q_\lambda\,,
\end{equation}
\noindent where $\delta_\sigma$ is an infinitesimal $U(1)$ transformation with parameter $\sigma$.  Using the explicit expression for $Q_\lambda$ in \eqref{Q}, we can compute the variation of the right hand side in \eqref{PB 1} and conclude
\begin{equation}
    \{ Q_\lambda, Q_\sigma \} = \hat{\alpha} \int_{\partial \Sigma} dx\, \lambda(x) \partial_x \sigma(x)\, ,
\end{equation}
\noindent which is the algebra of the chiral currents. By a calculation similar to the one above, we can show that analogous global symmetries are present in theories defined with the boundary condition \eqref{bc flat t x} with finite $\hat{\beta}$.


\providecommand{\href}[2]{#2}\begingroup\raggedright\endgroup

\end{document}